\documentclass[notitlepage, 
bibnotes,
prb,10pt, 
twocolumn,
superscriptaddress,
amsmath,amssymb,floatfix, aps]{revtex4-1}

\usepackage{enumitem}
\usepackage{amsfonts}
\usepackage{amsmath}
\usepackage{amssymb}
\usepackage{mathrsfs}
\usepackage{empheq}
\usepackage{tikz-cd}
\usepackage{graphicx}
\usepackage{dsfont}
\usepackage{dcolumn}
\usepackage{bm}
\usepackage{color,soul}
\usepackage{xcolor}
\usepackage{url}
\usepackage{tabularx}
\usepackage{array}
\usepackage{booktabs}
\usepackage{comment}
\usepackage{footnote}
\usepackage{lipsum}
\usepackage{mathrsfs} 
\usepackage[normalem]{ulem}
\usepackage{multirow}

\newcommand{\CORR}[1]{{\color{black} #1}}

\newcommand{\inlinecite}[1]{\citenum{#1}}

\renewcommand{\figurename}{\small{\bf Fig.}}

\makeatletter
\def\@caption@fignum@sep{\ \textbf{\textbar} \ }
\makeatother

\begin{document}

\title{Electric-current-assisted nucleation of zero-field hopfion rings}

\author{Xiaowen Chen}
\affiliation{Spin-X Institute, School of Physics and Optoelectronics, State Key Laboratory of Luminescent Materials and Devices, Guangdong-Hong Kong-Macao Joint Laboratory of Optoelectronic and Magnetic Functional Materials, South China University of Technology, 511442 Guangzhou, China
}%
\affiliation{Center for Electron Microscopy, South China University of Technology, 511442 Guangzhou, China}

\author{Dongsheng Song}
\affiliation{Information Materials and Intelligent Sensing Laboratory of Anhui Province, Institutes of Physical Science and Information Technology, Anhui University, 230601 Hefei , China}
\affiliation{Anhui Province Key Laboratory of Low-Energy Quantum Materials and Devices, High Magnetic Field Laboratory, HFIPS, Chinese Academy of Sciences, 230031 Hefei, China}

\author{Filipp N. Rybakov}
\email{philipp.rybakov@physics.uu.se}
\affiliation{Department of Physics and Astronomy, Uppsala University, SE-751 20 Uppsala, Sweden}

\author{Nikolai S. Kiselev}
\email{n.kiselev@fz-juelich.de}
\affiliation{Peter Gr\"{u}nberg Institute, Forschungszentrum J\"{u}lich and JARA, 52425 J\"{u}lich, Germany}

\author{Long Li}
\affiliation{Anhui Province Key Laboratory of Low-Energy Quantum Materials and Devices, High Magnetic Field Laboratory, HFIPS, Chinese Academy of Sciences, 230031 Hefei, China}       
\affiliation{University of Science and Technology of China, 230031 Hefei, China}

\author{Wen Shi}
\affiliation{Center for Electron Microscopy, South China University of Technology, 511442 Guangzhou, China}

\author{Rui Wu}
\affiliation{Spin-X Institute, School of Physics and Optoelectronics, State Key Laboratory of Luminescent Materials and Devices, Guangdong-Hong Kong-Macao Joint Laboratory of Optoelectronic and Magnetic Functional Materials, South China University of Technology, 511442 Guangzhou, China
}%

\author{Xuewen Fu}
\affiliation{Ultrafast Electron Microscopy Laboratory, The MOE Key Laboratory of Weak-Light Nonlinear Photonics, School of Physics, Nankai University, 300071 Tianjin, China}
\affiliation{School of Materials Science and Engineering, Smart Sensing Interdisciplinary Science Center, Nankai University, 300350 Tianjin, China}

\author{Olle Eriksson}
\affiliation{Department of Physics and Astronomy, Uppsala University, SE-751 20 Uppsala, Sweden}
\affiliation{Wallenberg Initiative Materials Science for Sustainability, Uppsala University, 75121 Uppsala, Sweden.}

\author{Stefan Bl\"{u}gel}
\affiliation{Peter Gr\"{u}nberg Institute, Forschungszentrum J\"{u}lich and JARA, 52425 J\"{u}lich, Germany}

\author{Rafal E. Dunin-Borkowski}
\affiliation{Ernst Ruska-Centre for Microscopy and Spectroscopy with Electrons, Forschungszentrum J\"{u}lich, 52425 J\"{u}lich, Germany}%

\author{Haifeng Du}
\affiliation{Information Materials and Intelligent Sensing Laboratory of Anhui Province, Institutes of Physical Science and Information Technology, Anhui University, 230601 Hefei , China}
\affiliation{Anhui Province Key Laboratory of Low-Energy Quantum Materials and Devices, High Magnetic Field Laboratory, HFIPS, Chinese Academy of Sciences, 230031 Hefei, China}

\author{Fengshan Zheng}
\email{zhengfs@scut.edu.cn}
\affiliation{Spin-X Institute, School of Physics and Optoelectronics, State Key Laboratory of Luminescent Materials and Devices, Guangdong-Hong Kong-Macao Joint Laboratory of Optoelectronic and Magnetic Functional Materials, South China University of Technology, 511442 Guangzhou, China
}
\affiliation{Center for Electron Microscopy, South China University of Technology, 511442 Guangzhou, China}%

\date{\today}

\maketitle

\textbf{
Magnetic hopfions are three-dimensional topological solitons — knotted, vortex-like spin configurations.
In chiral magnets, hopfions can appear as isolated structures or they can be linked to skyrmion strings.
Previous studies employed a sophisticated protocol and a special sample geometry to nucleate such hopfions linked to one or a few skyrmion strings.  
Here, we introduce an electric-current-assisted nucleation protocol that is simple and independent of the sample shape and size. 
The resulting hopfions exhibit extraordinary stability in the presence of both positive and negative magnetic fields, in perfect agreement with micromagnetic simulations.
We also present a comprehensive framework for classifying hopfions, skyrmions, and merons by deriving the corresponding homotopy group.
}

Topological magnetic solitons are localized, particle-like magnetization textures~\cite{Kovalev_90} that are characterized by well-defined sizes, positions, and velocities.
Their motion and interactions can be controlled by external stimuli like magnetic fields, temperature, or electric currents. 
This makes topological magnetic solitons highly promising for future applications~\cite{Fert_17, Tokura_20, Zhou_25}.

\begin{figure*}[ht]
    \centering
    \includegraphics[width=\linewidth]{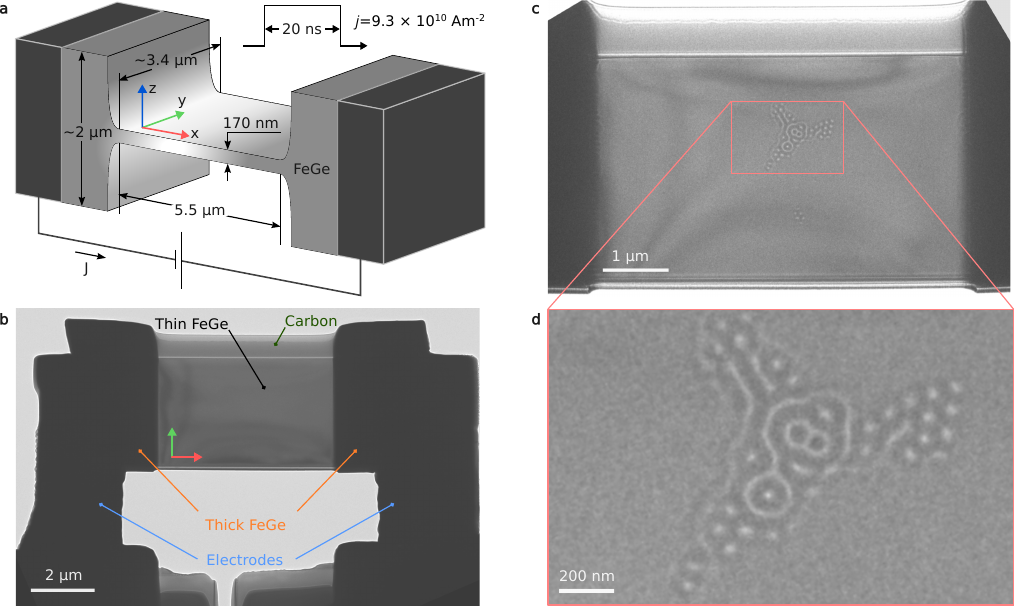}
    \caption{\textbf{Setup for electric-current-assisted experiments in the transmission electron microscope.} 
    \textbf{a}, Schematic representation of an FeGe sample.
    \textbf{b}, Low-magnification TEM image of the device viewed along the $+z$ direction. 
    Electrodes were connected to the thicker edges of the sample. 
    An amorphous, non-conductive carbon layer was deposited on one side of the sample during fabrication with focused ion beam (FIB) milling.
    \textbf{c}, Over-focus Lorentz TEM image of a representative magnetic state that appears after a short current pulse. The image is taken at a defocus distance of 700~$\mu$m and a temperature of 95~K.
    \textbf{d}, Magnified view of the area in \textbf{c}, showing a cluster of magnetic skyrmions and other magnetic textures.
    }
    \label{fig-setup}
\end{figure*}

Chiral magnets represent a unique class of crystals in which a wide variety of two-dimensional (2D) and three-dimensional (3D) topological solitons can coexist. 
Skyrmions are a representative example of 2D topological solitons.
In bulk crystals, skyrmions appear as string-like filamentary textures that penetrate the entire sample, but are essentially localized only in the 2D plane perpendicular to the string.
Purely 3D topological solitons, commonly referred to as hopfions, are magnetic textures that are well localized in all three spatial dimensions and, unlike skyrmions, can move freely in any spatial direction.

Topological solitons in classical field theory share some similarities with particles in quantum field theory. 
In particular, they are characterized by topological invariants that, to an extent, play the role of conserved quantum numbers -- quantities that remain unchanged even when solitons combine, interact or even merge. 
Remarkably, the topological invariants of solitons, together with their addition operation, define a homotopy group. 
In particular, the homotopy group of 3D topological solitons is closely related to Hopf's fundamental result on nontrivial mapping from the 3-sphere to the 2-sphere.
This connection gives rise to the term \textit{hopfion}, whose meaning varies throughout the literature and is also applied to phenomena that are beyond the scope of this work~\cite{Jiadong_18, Smalyukh_18, Sutcliffe_18, Kent_21, Yu_2023, Zhang_Du_2024}.

Currently, two classes of systems are known to host statically-stable magnetic hopfions: crystals with frustrated (competing) exchange interactions~\cite{Bogolubsky88, Sutcliffe_17, Liu_20, Rybakov_22, Klaui_23} and chiral magnetic crystals.
The latter is currently the only type of magnet where hopfions have been experimentally observed. These observations include hopfion rings linked to skyrmion strings~\cite{Zheng_23} and, more recently, isolated objects~\cite{Du2025Heliknoton, LaserHopfion}, which are usually referred to as \textit{heliknotons}~\cite{Tai_19}\CORR{$^{,}$\cite{Voinescu_20}}.

In the previous study, the protocol for hopfion ring nucleation in the transmission electron microscope (TEM) was technically sophisticated.
In particular, it required \textit{in~situ} monitoring of the magnetization texture across the entire sample, including the sample edges, as well as fine adjustment of the applied magnetic field based on the observed Lorentz TEM contrast.
Such an intricate approach depends on the shape, size, and quality of the sample and presents significant obstacles to the experimental study and practical applications of magnetic hopfions.

\begin{figure*}
    \centering
    \includegraphics[width=1\linewidth]{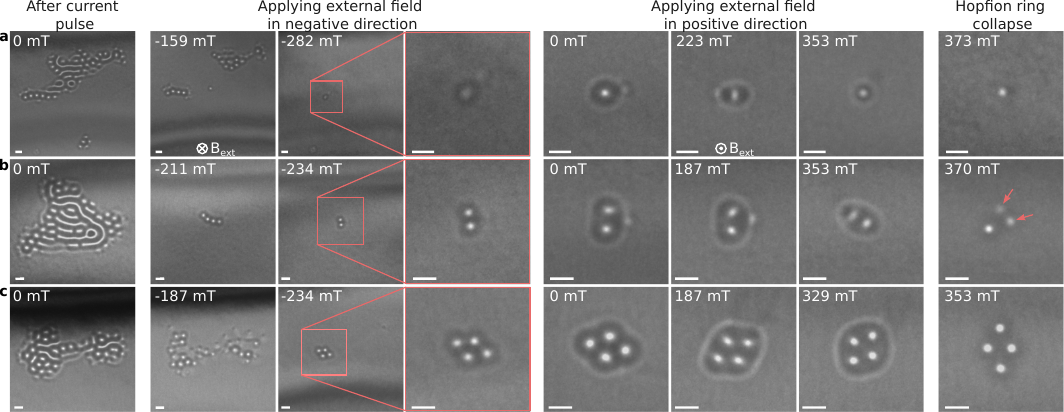}
    \caption{\textbf{Magnetic hopfion ring nucleation}.
    Each row represents a distinct sequence of over-focus Lorentz TEM images, illustrating the protocol for the nucleation of magnetic hopfion rings around (\textbf{a}) one, (\textbf{b}) two and (\textbf{c}) four magnetic skyrmion strings.
    The first column shows the initial configuration obtained after applying a short electric current pulse in zero magnetic field. 
    The second and third columns illustrate the evolution of the contrast under an external magnetic field applied in the negative direction (opposite to the initial direction of the saturating field). 
    This step is required to collapse large skyrmion clusters and to obtain an isolated hopfion ring.
    The fourth column provides a magnified view of the images in the third column. 
    The fifth to seventh columns show the evolution of Lorentz TEM contrast of hopfion rings as the applied field is increased in the positive direction. 
    The rightmost column displays the contrast immediately after the hopfion ring collapses.
    The arrows in \textbf{b} mark the positions of low-intensity spots corresponding to chiral bobbers or dipolar strings (torons).
    The images were recorded at a defocus distance of 700~$\mu$m and a temperature of 95~K.
    The scale bar in all images corresponds to 100~nm.
    }
    \label{fig2}
\end{figure*}

Here, we demonstrate that magnetic hopfions linked to skyrmion strings can be nucleated more efficiently using electric currents.
The experimental setup used in this study is similar to that utilized for the electric current generation of isolated hopfions in Ref.~\cite{Du2025Heliknoton}.
Building upon this approach, we report the remarkable stability of magnetic hopfion rings under both positive and negative applied magnetic fields and demonstrate excellent agreement between experimental phenomena and micromagnetic simulations. 
Furthermore, we present an in-depth homotopy group analysis that provides a comprehensive framework for classifying coexisting magnetic solitons (merons, skyrmions, and hopfions) in a background of helical modulations.

\textbf{Experimental setup.}
An electron-transparent sample for our experiment was fabricated from a single B20-type FeGe crystal using the focused ion beam (FIB) milling technique~\cite{Wang_22}.
The two ends of the sample were connected to thick Pt electrodes [Fig.~\ref{fig-setup}\textbf{a}]. 
The sample has lateral dimensions of 5.5~$\mu$m~$\times$~3.4~$\mu$m and a thickness of approximately 170~nm. 
The thickness of the sample in the electrode connection regions is significantly greater [Fig.~\ref{fig-setup}\textbf{b}].
A similar experimental setup was used earlier~\cite{Tang_21, Wang_22}.
Further details about the experimental setup and sample preparation are available in the Methods section.
An electric current pulse with a duration of 20~ns and a tunable voltage \CORR{amplitude} was applied to the sample.
The current density, $J$, was estimated to be $(9.3 \pm 0.2) \times 10^{10}$ A/m$^2$.
The experiment was carried out at a sample temperature of $\sim 95$~K.

\begin{figure*}[ht]
    \centering
    \includegraphics[width=1\linewidth]{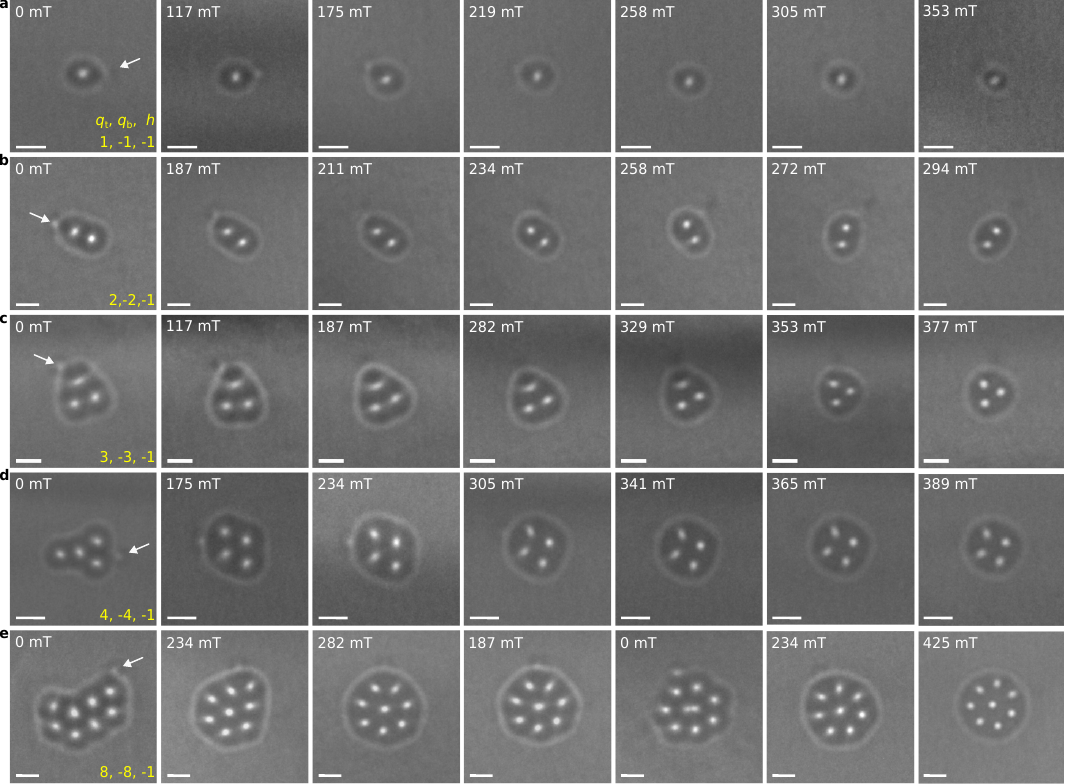}
    \caption{\textbf{Magnetic-field-driven evolution of hopfion rings}.
    Each row shows a sequence of over-focus Lorentz TEM images recorded between zero magnetic field and the field at which the hopfion ring collapses.  
    The strength of the magnetic field is labeled in the upper left corner of each image.
    In \textbf{a}-\textbf{d}, the magnetic field increases from left to right. 
    In \textbf{e}, the magnetic field increases, decreases to zero and then increases again.
    The arrows in the images recorded in zero magnetic field indicate the position of a bump that is present in each hopfion ring.
    With increasing field, such bumps become less pronounced.
    All of the images were recorded at a sample temperature of 95~K and a defocus distance of 700~$\mu$m.
    The scale bar in all images is 100~nm.
    The three indices shown in the lower right corner in the first column represent the topological charges of the magnetic textures in the corresponding row: two 2D topological indices $q_t$ and $q_b$ and one 3D topological index $h$.
    The cone winding number is $v=2$ for all configurations.
    See Extended Data Fig.~\ref{fig:SS5} for corresponding simulated images.
    }
    \label{fig3}
\end{figure*}

\textbf{Current-assisted nucleation of hopfions.}
Figure~\ref{fig-setup}\textbf{c} shows a representative magnetic state after applying current pulses in zero magnetic field.
Before applying the current, the sample was saturated by a strong positive perpendicular magnetic field.
(See Extended Data Video~1 for sequential switching of the magnetic state under 1~Hz electric current pulses).
Based on previous studies, the appearance of a cluster of skyrmions and helical spirals results from Joule heating induced by the electric current~\cite{Yu_20,Qin_22,Zhao_22}.
\CORR{While other effects, such as spin-transfer torque and Oersted fields, may also contribute, they are not expected to play a leading role here.}
\CORR{The nucleation of hopfion rings is a stochastic process.
In this work, we estimate a probability of approximately 10\% for the formation of hopfion rings at a pulse current density of $9.3 \times 10^{10}$~A/m$^2$.
}

The absence of magnetic contrast around the cluster in Fig.~\ref{fig-setup}\textbf{c} and \textbf{d} suggests that most of the sample volume is occupied by helical or conical modulations with a $\mathbf{k}$-vector perpendicular to the plate.
In zero magnetic field, the ground state of the FeGe thin film is a helical spiral with a $\mathbf{k}$-vector in the plane of the film~\cite{Yu_11, Du_18}, whereas phases with other $\mathbf{k}$-vectors appear as metastable states. 
This effect becomes pronounced in extended plates and at low sample temperatures.

\begin{figure*}
    \centering
    \includegraphics[width=\linewidth]{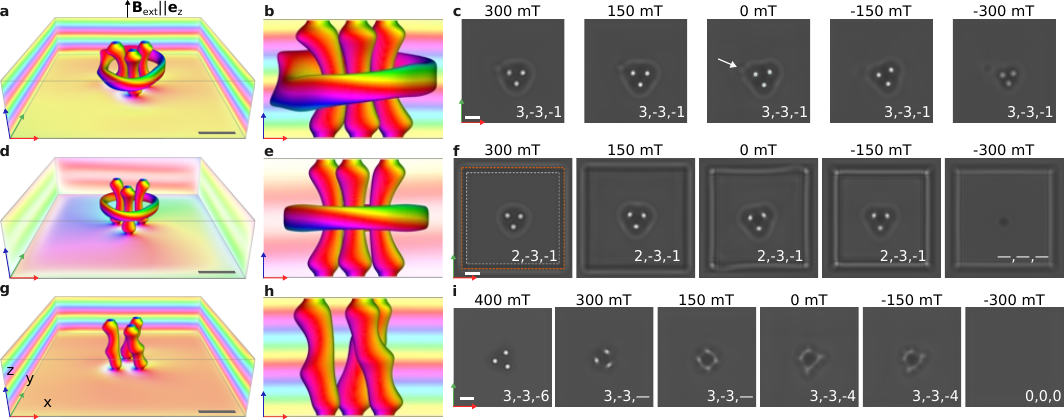}
    \caption{\textbf{Micromagnetic simulations for a 170-nm-thick FeGe plate.}
    \textbf{a}, 
    Hopfion surrounding the three skyrmion strings in an external magnetic field of 300~mT, obtained through micromagnetic energy minimization using parameters specific to FeGe. 
    The calculation was performed in a 640~nm~$\times$~640~nm~$\times$~170~nm domain with periodic boundary conditions in the $x$-plane and a 7.5~nm damaged surface layer on the top and bottom surfaces. 
    The magnetization is visualized by an isosurface for $m = 0$ and at the domain edges.
    Directions are color-coded using the standard scheme: white and black pixels indicate moments aligned parallel and antiparallel to the $z$-axis, while red, green and blue represent the azimuthal angle relative to the $x$-axis.  
    \textbf{b}, Side view along the $y$-axis of the configuration shown in \textbf{a}.  
    \textbf{c}, Magnetic field evolution of the simulated over-focus Lorentz TEM contrast for the configuration shown in \textbf{a} in the presence of external magnetic fields applied along the positive and negative $z$-axis directions. 
    The arrow in the middle image (0~mT) marks a bump similar to that observed in the experimental images in Fig.~\ref{fig2} and \ref{fig3}.
    The three indices in the lower right corner indicate the topological charge, which consists of two 2D topological indices  $q_t$ and $q_b$ and one 3D topological index $h$. 
    For all of the textures in this figure, the background winding number is $v=2$.
    \textbf{d}, Stable magnetization configuration, similar to that in \textbf{a}, but calculated with open boundary conditions at 300~mT.  
    \textbf{e}, Side view of the isosurface for $m = 0$ showing greater symmetry of the hopfion, as compared to the state shown in \textbf{b}  
    \textbf{f}, Simulated Lorentz TEM images similar to those in \textbf{c}, but for the confined geometry depicted in \textbf{d}.
    The outer (red) dashed square marks the shape of the sample.
    The inner (white) dashed square marks the volume in which the topological indices were calculated to avoid the impact of edge modulations.
    At approximately $-$250~mT, the edge modulations contract towards the center, leading to the collapse of the hopfion ring and skyrmions.
    The configuration in \textbf{f} at $-$300~mT contains singular points, and topological indices are undefined.
    \textbf{g}, Skyrmion cluster composed of three skyrmion strings at 300~mT.  
    \textbf{h}, Side view of the skyrmion cluster depicted in \textbf{g}.  
    \textbf{i}, Simulated Lorentz TEM images of the skyrmion cluster, showing a braiding effect with decreasing applied magnetic field.
    At approximately $-$200~mT, the skyrmions in the cluster collapse.
    The scale bar in all images is 100~nm.
    }
    \label{fig4}
\end{figure*}

In order to isolate magnetic hopfion rings from the current-induced clusters, we applied an external magnetic field perpendicular to the plate.
Representative examples of magnetic-field-driven nucleation of the hopfion ring are shown in Fig.~\ref{fig2}. 
Each row shows a distinct experimental series of four stages arranged in four columns. 
The first column shows initial skyrmion clusters in zero applied magnetic field after applying an electric current pulse.
After cluster nucleation, we applied a magnetic field in the negative direction, \emph{i.e.}, opposite to the direction of the magnetic field that was initially applied to saturate the sample.
In an opposite magnetic field of $\sim 230 - 280$~mT (see the middle column of Fig.~\ref{fig2}), most of the skyrmions in the cluster collapse and only configurations with typical hopfion ring contrast remain stable.
We then reduced the applied magnetic field to zero, inverted its direction and increased the field in the positive direction.
Remarkably, the sequential application of a moderate magnetic field in the positive and negative directions, or switching the field off, did not result in instability of the hopfion rings (see, \emph{e.g.}, Fig.~\ref{fig3}\textbf{e} and Extended Data Fig.~\ref{fig:S0} and~\ref{fig:S8}).

As illustrated in the last column of Fig.~\ref{fig2}, the hopfion rings collapse only when the magnetic field approaches $\sim 350-370$~mT. 
In some cases, collapse of the hopfion ring leads to collapse of the skyrmion strings as shown in Fig.~\ref{fig2}\textbf{b}.
When the hopfion ring collapses, we often observe the appearance of spots with low-intensity contrast, which are associated with chiral bobbers\cite{Zheng_18} or dipole strings~\cite{Muller_20, Savchenko_22, Kuchkin_25}. (See Fig.~\ref{fig2}\textbf{b} and Extended Data Fig.~\ref{fig:S3}\textbf{h}).
However, in most cases the hopfion rings collapse, leaving no visible contrast (see Fig.~\ref{fig2}\textbf{a} and \textbf{c}).
This phenomenon was also reported earlier~\cite {Zheng_23}.

In order to demonstrate the efficiency of electric-current-assisted hopfion ring nucleation, Fig.~\ref{fig3} shows a collection of hopfion rings with varying numbers of skyrmion strings and their field evolution.
The images illustrate the stabilty of hopfion rings in the presence of relatively high magnetic fields, which sometimes exceed $400$~mT.
Extended Data Fig.~\ref{fig:S1} and ~\ref{fig:S6} show the entire field of view corresponding to the configurations in Fig.~\ref{fig3}\textbf{a} and \textbf{c}, respectively.
These results also highlight the translational and rotational mobility of a hopfion ring within the sample during field evolution.
\CORR{
In our high-quality sample, we did not observe either the hopfion pinning effect or a predominant area of hopfion ring nucleation under the current pulses.
}

Increasing the applied magnetic field weakens the intensity of the hopfion rings and reduces the distance between the skyrmions, as reported in the previous study~\cite{Zheng_23}. 
Furthermore, the skyrmion clusters become more symmetric with increasing field, while the surrounding hopfion rings become more circular.
Extended Data Fig.~\ref{fig:S3} shows a series of additional experimental images that illustrate the field evolution of a hopfion ring enclosing three skyrmion strings.
Extended Data Fig.~\ref{fig:SS5} shows simulated Lorentz TEM images, which are in  perfect agreement with the experimental images shown in Fig.~\ref{fig3}.

\begin{figure*}
    \centering
    \includegraphics[width=\linewidth]{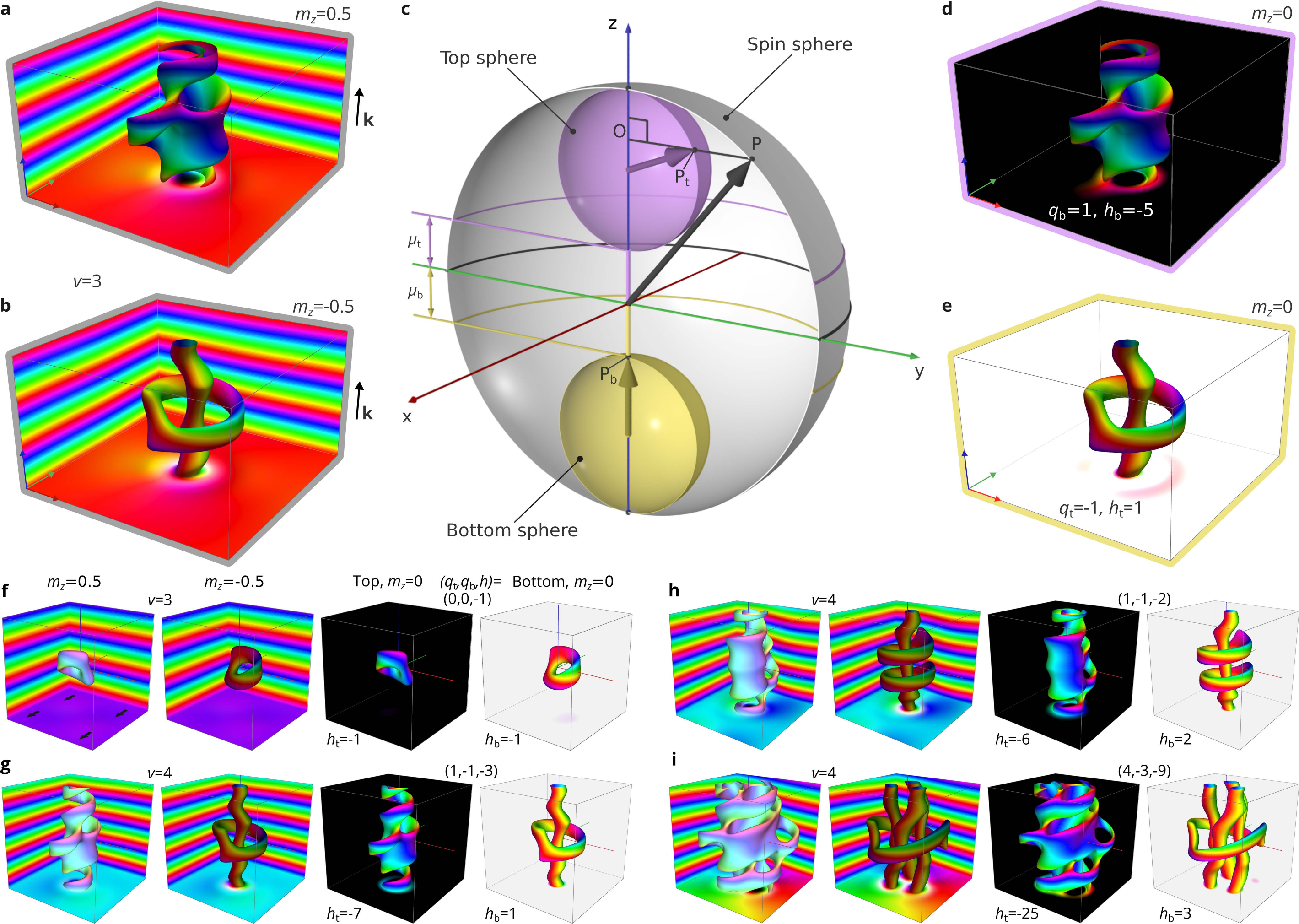}
    \caption{
    \textbf{Dumbbell map transformation for magnetic textures with topological index $\bm{(q_\text{t}=-1,q_\text{b}=1,h=-2)}$.}
    \textbf{a} and \textbf{b}, Visualisations of magnetization textures obtained by energy minimization in a domain of $5L_\text{D}\times 5L_\text{D}\times 3L_\text{D}$ with parameters for FeGe (see the Methods section for details), periodic boundary conditions in all three directions and isosurfaces of $m_z=0.5$ and $m_z=-0.5$, respectively. 
    The winding number $v=3$ indicates the number of helical modulations along the $z$-axis.
    \textbf{c}, Dumbbell projection from the spin sphere onto a dumbbell composed of top and bottom spheres connected by a handle segment.
    The map is parameterized by the shape parameters $\mu_\mathrm{t}$, $\mu_\mathrm{b}$ and the middle plane position between the spheres $\mu_0 = \cos \Theta_\mathrm{c}$.  
    The black arrow represents an example of a spin being projected. 
    Its projections onto the top and bottom spheres are shown as magenta and yellow arrows, respectively.
    Extended Data Video~1 illustrates this mapping for varying parameters  $\mu_\mathrm{t}$, $\mu_\mathrm{b}$ and $\mu_0$.
    \textbf{d} and \textbf{e}, Vector fields corresponding to the normalized projection into the top and bottom spheres, respectively. 
    The indices $q_\mathrm{t}$ and $q_\mathrm{b}$ denote the 2D topological index for the top and bottom projections.
    Similarly, $h_\mathrm{t}$ and $h_\mathrm{b}$ indicate the 3D topological index for the corresponding textures.
    \CORR{
    \textbf{f-i}, Collection of topological magnetic textures with different topological index $(q_\mathrm{t}, q_\mathrm{b}, h)$, where $h=h_\mathrm{t}+vq_\mathrm{t}=h_\mathrm{b}+vq_\mathrm{b}$.
    Each image depicts a statically stable magnetic configuration, obtained by energy minimization of the micromagnetic functional with FeGe parameters at zero magnetic field under periodic boundary conditions (Methods).  
    Each image consists of four panels in a single row.
    Each panel visualizes the magnetization at the edges and the specific vector field isosurface. 
    The first and second panels show isosurfaces for $m_z = 0.5$ and $m_z = -0.5$, respectively.
    The third and fourth panels show projections of magnetization on the top and bottom spheres of the dumbbell, respectively.  
    The winding number $v$ between the first and second column indicates the number of helical modulations along the $z$-axis.
    The triplet of indices $(q_\mathrm{t}, q_\mathrm{b}, h)$ between the third and fourth panels represents the topological charge of the corresponding configuration.
    \textbf{f}, Isolated hopfions embedded into a helix (heliknoton).
    \textbf{g}, Hopfion ring configuration, similar to the state shown in panels \textbf{a} and \textbf{b}, but with a winding number, $v=4$.
    \textbf{h}, Double hopfion rings on a single skyrmion string.  
    \textbf{i}, Single hopfion ring on a skyrmion braid composed of three strings.  
    }    
    }
    \label{fig5}
\end{figure*}

\textbf{Stability and characteristic features of hopfion rings.}
A feature that appears in most of the experimental images is a bump in each hopfion ring, which is most evident in images recorded in low applied magnetic fields (see, \emph{e.g.}, Fig.~\ref{fig3}, marked by white arrows).
Extended Data Fig.~\ref{fig:S3} illustrates the varying position of the bump and the rotational degree of freedom of the hopfion ring during applied magnetic field cycles.
Extended Data Fig.~\ref{fig:S4} contains an additional collection of images recorded in both over-focus and under-focus regimes, which show that the bumps become less visible with increasing applied magnetic field.

Micromagnetic simulations were performed to explain the origin of the bumps. (See the Methods section for details).
Figure~\ref{fig4}\textbf{a} shows the equilibrium configuration of a hopfion ring surrounding three skyrmion strings.
Simulations were performed with periodic boundary conditions in the $xy$-plane to mimic the extended plate.
The configuration is embedded in a regular conical phase, which distorts the toroidal shape of the hopfion ring.
The distortion is visible in the side view shown in Fig.~\ref{fig4}\textbf{b}.
The simulated Lorentz TEM images in Fig.~\ref{fig4}\textbf{c} show the same bumps as in the experimental images.
In the case of open boundary conditions [Fig.~\ref{fig4}\textbf{d}], the surrounding conical phase contains a vortex (meron) with a core at the center of the domain.
In this scenario, the hopfion ring appears more symmetrical [Fig.~\ref{fig4}\textbf{e}].
The corresponding Lorentz TEM images are free of the characteristic bump [Fig.~\ref{fig4}\textbf{f}].
The presence of the bumps can be explained by a slight distortion of the hopfion ring embedded in a regular conical phase.
\CORR{
In particular, the step-like bending of the hopfion ring isosurface, visible in Fig.~4\textbf{a} and \textbf{b}, leads to a localized deviation of the projected in-plane magnetization from the symmetric case.
This deviation coincides spatially with the position of the bump observed in the corresponding Lorentz TEM images.
}
The presence or absence of this feature in Lorentz TEM contrast, therefore, depends on the lateral size of the sample.

The images in Fig.~\ref{fig4}\textbf{c} and \textbf{f} also show that the sizes of the hopfion rings and the magnetic field range for their existence in the extended film and the sample of finite size are different.
Demagnetizing field effects and the presence of edge modulations result in hopfion ring collapse in a sample of finite size at negative fields.
In the case of an extended film, the hopfion ring remains stable in the range from $+$300 to $-$300~mT [Fig.~\ref{fig4}\textbf{c}].
In contrast, the hopfion ring collapses at about $-$250~mT in the confined sample [Fig.~\ref{fig4}\textbf{f}].
The edge modulations contract toward the center of the sample, leading to the collapse of the hopfion ring and skyrmions and the appearance of a state containing Bloch points.
Similar configurations have been considered in Refs.~\cite{Muller_20, Garst_22, Kuchkin_25}, see also the final image in Fig.~\ref{fig4}\textbf{f} and Extended Data Fig.~\ref{fig:E2}\textbf{j} and \textbf{l}.
Contraction of the edge modulations, leading to collapse of the internal configurations, explains why stability of the hopfion rings up to strong negative fields was not observed in the previous study~\cite{Zheng_23}, in which a relatively small square-sized sample ($\sim$1~$\mu$m~$\times$~1~$\mu$m) was studied.

The protocol presented above and micromagnetic simulations (Fig.~\ref{fig4}\textbf{d}-\textbf{f}) suggest that skyrmions linked to hopfions exhibit stability in a wide range of positive and negative magnetic fields.
In contrast, clusters of isolated skyrmions exhibit much weaker stability in negative fields.
The results of the micromagnetic simulations are in perfect agreement with the experimental observations.
Figure~\ref{fig4}\textbf{g}-\textbf{i} illustrates the behavior of a stand-alone cluster of three skyrmions in positive and negative perpendicular magnetic fields.  
Skyrmions form almost straight strings in a strong positive field of $400$~mT.  
As the field is decreased, the skyrmion cluster transforms into skyrmion braids.
The latter behavior is evident from the characteristic blurry contrast of skyrmion strings in Lorentz TEM images~\cite{Zheng_21}. 
The skyrmion braids remain stable in zero field and survive in moderately negative fields.
However, at approximately \CORR{$-200$}~mT, all of the skyrmions in the cluster collapse. 
\CORR{An} isolated skyrmion exhibit\CORR{s} similar behavior in negative fields~\cite{Leonov_18}. 
\CORR{Thus, skyrmions inside hopfion ring are more stable against negative magnetic fields than isolated skyrmions (compare Fig.~\ref{fig4}\textbf{c} and \textbf{i}).
The surrounding hopfion ring suppresses the expansion of the skyrmion core and prevents its transformation into a helical spiral state, thereby extending the stability range of the enclosed skyrmions under negative magnetic fields.}

In addition to the shape and size of the sample, an essential factor underlying the phenomena presented here is the presence of a damaged layer on the surface of the sample, resulting from sample preparation using the FIB milling technique\cite{Wolf_22, Savchenko_22}.
Previous studies have shown that this FIB-damaged layer can be approximated as amorphous FeGe, which exhibits the same magnetization and exchange stiffness as B20-type FeGe, but has negligibly weak DMI~\cite{Savchenko_22,Zheng_23}.
Extended Data Fig.~\ref{fig:S7} highlights the important role of the damaged layer by comparing the field evolution of a hopfion ring in systems with and without it.

\CORR{
It is important to emphasize that neither the sample shape nor the presence of the FIB-damaged surface layer is responsible for stabilizing hopfion rings.
The stability of hopfion rings is governed primarily by the competition between the isotropic Heisenberg exchange and the bulk-type DMI.
Hopfion rings studied here occur in a helical magnetic background with zero net magnetization.
As a result, their lowest-energy configuration is realized at zero applied magnetic field, and any external magnetic field increases their energy, leading to a shrinking and eventual collapse of the hopfion rings.
This behavior differs qualitatively from hopfions stabilized in chiral magnets but embedded in a field-polarized ferromagnetic background, where stability is achieved through geometrical confinement and specially imposed boundary conditions, as discussed in Refs.
~\cite{Tai18PNAS, Ackerman17, Smalyukh_18, Jiadong_18, Sutcliffe_18}.
}

\textbf{Homotopy group analysis.}
The stability of hopfion rings in an invertible magnetic field raises an important question concerning the topological classification of such magnetic configurations.
In a previous study~\cite{Zheng_23}, we demonstrated that hopfion rings can be classified using the skyrmion–hopfion topological charge, represented as an ordered pair of integers.
This classification applies to magnetic configurations in a sufficiently polarized background.
An example is the conical phase with a $\mathbf{k}$-vector directed along the field. 
In this scenario, the magnetization vectors of the conical phase form a circle on the $\mathbb{S}^2$ sphere, which can always be contracted continuously to a single point $\mathbf{m}_0$, the closest equidistant point to any point on that circle. 
However, in the present experiment, the background phases are not necessarily field-polarized.
In particular, in zero magnetic field the conical phase transforms into a helical phase, which forms a circle corresponding to the equator of the $\mathbb{S}^2$ sphere.
Such a state has no net magnetization.
In this case, the classification approach proposed in Ref.~\cite{Zheng_23} becomes ambiguous due to the absence of a well-defined polarity.
In order to generalize the classification of topological magnetic textures, we considered an additional limiting case, in which the vacuum is treated as noncontractible to a single point on the spin sphere.
(See details of the homotopy analysis and the derivation of the corresponding group in the Methods section).
Here, we focus on the practical implications, which are now discussed in detail.

In order to calculate the topological charge of the configuration embedded in the cone phase or helix, we use an auxiliary map, as illustrated in Fig.~\ref{fig5}.
We refer to this map, which was introduced in Ref.~\cite{VortMerSkyrm}, as the \textit{dumbbell} map, as a result of the shape of the target space, which consists of two spheres that are connected by a segment (handle).
Figure~\ref{fig5} illustrates the application of the dumbbell map to a representative magnetic configuration of a hopfion ring on a single skyrmion string embedded in a helix, with the $\mathbf{k}$-vector parallel to the $z$-axis.
Figure~\ref{fig5}\textbf{a} and \textbf{b} illustrate the same magnetic configurations obtained by micromagnetic energy minimization in zero magnetic field in a domain with periodic boundary conditions in all directions. (See the Methods section).
In Fig.~\ref{fig5}\textbf{a}, the magnetic texture is visualized in the form of an $m_z=0.5$ isosurface, while in Fig.~\ref{fig5}\textbf{b} an $m_z=-0.5$ isosurface is shown for the same configuration.

Application of the dumbbell map depicted schematically in Fig.~\ref{fig5}\textbf{c} results in two auxiliary spin configurations, which are referred to as the top projection and the bottom projection, as shown in Fig.~\ref{fig5}\textbf{d} and \textbf{e}, respectively. 
The map ensures that both configurations are embedded in a ferromagnetic background with $\mathbf{m}_0=(0,0,-1)$ for the top projection and $\mathbf{m}_0=(0,0,+1)$ for the bottom projection.
For each of the two projections, one can compute the 2D topological indices $q_\mathrm{t}$ and $q_\mathrm{b}$ and the 3D topological indices $h_\mathrm{t}$ and $h_\mathrm{b}$,
where `t' and `b' refer to the top and bottom projections. (See the Methods section for details). 
Our analysis reveals that, among the four indices, only three are linearly independent.
Consequently, the corresponding topological invariants are represented by an ordered triple of integers $(q_\mathrm{t}, q_\mathrm{b}, h)$ and form a homotopy group that is isomorphic to $\mathbb{Z}^3$. (See the Methods section for details).
Configurations with identical topological charges -- identical tuples of topological indices -- are topologically equivalent.  
Such an equivalence is meaningful only for configurations that are embedded in identical vacuums, meaning that they share the same magnetization background.
In this study, we focus on magnetic textures that are embedded in a phase with modulations along the $z$-axis. 
The diversity of magnetic configurations embedded in such helical or conical backgrounds can be considered as classes indexed by $v$ -- the number of periods along the $z$-axis.
It means that such magnetic configurations can be a distinct part of a larger, or even an infinite, texture (\emph{e.g.}, in the case of periodic boundary conditions).
Since $v$ inherently represents a winding number of the vacuum state along the $z$-axis, the sign of $v$ depends on the chirality of the helical modulations. (See, \emph{e.g.}, \CORR{Fig.~\ref{fig5}\textbf{f} and} Extended Data Fig.~\ref{fig:E1}\textbf{a}\CORR{,} and \CORR{Fig.~\ref{fig5}\textbf{g} and} \CORR{Extended Data Fig.~\ref{fig:E1}\textbf{b}}).
The topological index $h$ is related to the winding number $v$ and the in-plane 2D topological indices $q_t$ and $q_b$ as follows: $h = h_\mathrm{t} + v\,q_\mathrm{t}=h_\mathrm{b} + v\,q_\mathrm{b}$.
It should be noted that none of the above indices correspond directly to the \textit{pure} skyrmion or hopfion numbers ($Q$ or $H$), as they belong to different homotopy groups and describe distinct topological properties.
However, the pure skyrmion and hopfions can be recovered if the magnetic configuration is continuously transformed into a state with a uniformly polarized background.
In that case, $Q$ and $H$ depend on the polarity of the background magnetization (see the Methods section).

The configuration depicted in Fig.~\ref{fig5}\CORR{\textbf{a-e}} has a winding number of $v=3$ and a topological index of $(q_\mathrm{t}, q_\mathrm{b}, h) = (1,-1,-2)$.
Topological indices $(q_\mathrm{t}, q_\mathrm{b}, h)$ are also provided for the magnetic textures shown in Figs.~\ref{fig3} and~\ref{fig4}.
Note the difference between the $q_t$ index in Fig.~\ref{fig4}\textbf{c} and \textbf{f} due to the different background states.
The topological indices of the hopfion rings in the extended plate (Fig.~\ref{fig4}\textbf{a}-\textbf{c}) and in the confined sample (Fig.~\ref{fig4}\textbf{d}-\textbf{e}) remain fixed over the entire range of stability fields.
In contrast, the $h$ index of the free-standing skyrmion cluster (Fig.~\ref{fig4}\textbf{i}) varies with the strength of the applied magnetic field, indicating a topological distinction between a straight skyrmion string and a skyrmion braid.
This behavior reflects the fact that the 3D topological index $h$ is not equivalent to the hopfion number, but has a more general meaning.
Other hopfion-free configurations, such as skyrmions and vortices embedded in the cone phase~\cite{Rybakov_2015, Garst_22, tz4v-s5m5}, naturally fit within the unified classification presented here and exhibit a non-zero $h$ index.
As indicated in Fig.~\ref{fig4}\textbf{i} for applied magnetic fields of 300 and 150~mT, the $h$ index may not be well-defined at intermediate magnetic fields, when the system is in a transient state.

Other representative configurations with nontrivial topology are shown in \CORR{Fig.~\ref{fig5}\textbf{f-i} and} Extended Data Fig.~\ref{fig:E1}.
In particular, \CORR{Fig.~\ref{fig5}\textbf{f} and} Extended Data Fig.~\ref{fig:E1}\textbf{a} illustrate an isolated magnetic hopfion embedded in a helix (heliknoton)~\cite{Voinescu_20, LaserHopfion}. 
Note that the systems in \CORR{Fig.~\ref{fig5}\textbf{f}} and \CORR{Extended Data Fig.~\ref{fig:E1}\textbf{a}} have opposite chiralities and opposite signs of the 3D topological index.
The configurations in \CORR{Fig.~\ref{fig5}\textbf{g}} and \CORR{Extended Data Fig.~\ref{fig:E1}\textbf{b}} show a hopfion linked to a single skyrmion string, similar to the state shown in Fig.~\ref{fig5} but in a larger simulated domain that accommodates helical spirals with winding number $v=4$. 
Note that the configurations in \CORR{Fig.~\ref{fig5}\textbf{g}} and \CORR{Extended Data Fig.~\ref{fig:E1}\textbf{b}} also have opposite chiralities. 
The configurations in \CORR{Fig.~\ref{fig5}\textbf{h}} and \CORR{Extended Data Fig.~\ref{fig:E1}\textbf{c}} illustrate different linkages between hopfions and skyrmion strings.
The state depicted in \CORR{Fig.~\ref{fig5}\textbf{i}} presents one of the most complex configurations, consisting of a skyrmion braid linked to a single hopfion ring embedded in a vortex cone background.
Extended Data Fig.~\ref{fig:E1}\CORR{\textbf{d}} shows a so-called hybrid skyrmion string, as studied previously in Ref.~\cite{Kuchkin22}. 
The contributions of the vacuum and the skyrmion braiding effect to the topological charge of the system are illustrated in Extended Data Figs~\ref{fig:E2} and~\ref{fig:E3}, respectively.
The diverse set of topological configurations illustrated in \CORR{Fig.~\ref{fig5}\textbf{f-i},} Extended Data Figs~\ref{fig:E2},~\ref{fig:E1} and \ref{fig:E3} demonstrate that the topological classification presented here provides the most comprehensive framework for classifying 2D and 3D topological magnetic solitons, including merons, skyrmions and hopfions.

\textbf{Conclusions.} 
We have successfully realized the nucleation of hopfion rings in an FeGe film with large lateral dimensions without the need for geometric confinement and their efficient generation by electric currents.
The successful nucleation of hopfions \emph{via} energy injection from pulsed electric currents suggests that other ultrafast excitations, such as laser or field pulses, may also serve as promising approaches for hopfion generation.
Our results open new pathways for studying hopfion dynamics under external stimuli and pave the way for potential applications in spintronics and unconventional computing.
The hopfions that we observe are stable in zero magnetic field, facilitating future tomographic studies of such three-dimensional magnetic configurations using high-resolution transmission electron microscopy~\cite{Midgley_09}\CORR{$^{,}$\cite{Yu_22}} and X-ray microscopy~\cite{donnelly-x-ray}.

\section*{Methods}
\textbf{Specimen preparation.}
An FeGe TEM specimen, fixed on an \emph{in~situ} electrical chip, was prepared from a single crystal of B20-type FeGe using a standard lift-out method from the bulk using a focused ion beam instrument (Helios Nanolab 600i, FEI)~\cite{Wang_22}.

\textbf{Electric current experimental setup and TEM magnetic imaging.}
Fresnel magnetic images were recorded with a TEM instrument (Talos F200X, FEI), operated at 200 kV in Lorentz mode. A single-tilt liquid-nitrogen specimen holder (Model 616.6 cryotransfer holder, Gatan) was used in a temperature range of 95 to 300 K. The experiments described in the main text were all performed at 95 K. The perpendicular magnetic field was controlled by adjusting the current of the objective lens. The electric current pulses were supplied by a voltage source (AVR-E3-B-PN-AC22, Avtech Electrosystems), with a pulse width of 20 ns and a frequency of 1 Hz~\cite{Tang_21}. \CORR{The applied voltage amplitude was tunable and set to 43.5 V in the experiment. The device resistance at 95 K was approximately 500 $\Omega$. The current pulses had a square waveform with pulse rise time/overshoot of 1-2 ns.}

\textbf{Micromagnetic simulations.}
The micromagnetic model of isotropic chiral magnet includes four energy terms: the Heisenberg exchange, the Dzyaloshinskii-Moriya interaction (DMI), the Zeeman energy, and the demagnetizing field energy term: 
\begin{align} 
\mathcal{E}\!=\!\int\limits_{V_\mathrm{m}}\!d\mathbf{r}\ 
&\mathcal{A}\sum\limits_{i=x,y,z} |\nabla m_i|^2 
+\mathcal{D}\,\mathbf{m}\!\cdot(\nabla\!\times\!\mathbf{m})
- M_\mathrm{s}\,\mathbf{m}\!\cdot\!\mathbf{B} + \nonumber\\
&+ \frac{1}{2\mu_0}\int\limits_{\mathbb{R}^3}\!d\mathbf{r}\ 
\sum\limits_{i=x,y,z}|\nabla A_{\mathrm{d}, i}|^2~,
\label{Ham_m}
\end{align}
where 
${\mathbf{m}(\mathbf{r})}=\mathbf{M}(\mathbf{r})/M_\text{s}$ is a unit vector field that defines the direction of magnetization,  
$M_\text{s}=|\mathbf{M}(\mathbf{r})|$ is the saturation magnetization,
$\mathbf{B} = \mathbf{B}_\text{ext} + \nabla\!\times\!\mathbf{A}_\text{d}$ is the total magnetic field,
${\mathbf{A}_\text{d}(\mathbf{r})}$ is the magnetization induced magnetic vector potential,
$\mathcal{A}$ is the Heisenberg exchange stiffness constant, 
$\mathcal{D}$ is the constant of isotropic bulk-type DMI and $\mu_0$ is the vacuum permeability.
All simulations presented in this work were performed using FeGe material parameters that have been well tested in previous studies~\cite{Zheng_18, Zheng_21, Savchenko_22,Zheng_23}:
$\mathcal{A}=4.75$~pJm$^{-1}$, $\mathcal{D}=0.853$~mJm$^{-2}$ and $M_\text{s}=384$~kAm$^{-1}$. 
For these parameters, the equilibrium period of the helical spin spiral is given by $L_\mathrm{D}=4\pi\mathcal{A}/\mathcal{D}=70$ nm, and the critical field for the cone phase saturation is $B_\mathrm{c}=B_\mathrm{D}+\mu_0 M_\mathrm{s}=0.682$ T, where $B_\mathrm{D}=\mathcal{D}/(2M_\mathrm{s}\mathcal{A})=0.199$ T. 

The equilibrium solutions of the Hamiltonian~(\ref{Ham_m}) were obtained via numerical energy minimization using the Excalibur code~\cite{Excalibur} and Mumax\cite{mumax}.
Theoretical Lorentz TEM images were computed following the method described in Ref.~\cite{Zheng_21}, which is implemented in Excalibur.
The FIB-damaged layer was modeled by setting $\mathcal{D} = 0$, following the approach in Ref.~\cite{Savchenko_22}.
For the simulations shown in Fig.~\ref{fig4}, we used a computational domain of $L_x=L_y=640$ nm, $L_z=170$ nm, with a finite-difference mesh of $256 \times 256 \times 68$ nodes.
For the calculations presented in Fig.~\ref{fig5} and Extended Data Figs.~\ref{fig:E2},~\ref{fig:E1}, and~\ref{fig:E3}, we used different domain sizes but the same mesh density, periodic boundary conditions, and neglected demagnetizing fields.

\textbf{Algebraic topological analysis.} 
We classify complex magnetic textures, consisting of skyrmion/meron strings and hopfions, according to their homotopy types. 
Each homotopy type~\cite{Hatcher} is a class of textures that can be continuously transformed into each other and is quantified by topological invariants.
When identifying topological invariants, we consider realistic boundary conditions, taking into account that magnetic textures are generally localized against the background of non-perfect magnetic phases, such as near-collinear ferromagnetic or distorted helical/conical states. 
To address this, we employ a specialized technique -- the auxiliary dumbbell map~\cite{VortMerSkyrm}, which establishes a connection to the category of pointed spaces~\cite{Hatcher, ModernClassicalHomotopyTheory}.

\textbf{Auxiliary dumbbell map.} 
The dumbbell map is schematically illustrated in Fig.~\ref{fig5}\textbf{c} and Supplementary Video~1. 
It maps every spin from the~$\mathbb{S}^2$ sphere onto a space composed of two spheres (top and bottom) joined by a handle -- a line segment connecting the south pole of the top sphere to the north pole of the bottom sphere. 
For example, in Fig.~\ref{fig5}\textbf{c}, the point P on the spin sphere is projected onto P$_\mathrm{t}$ on the top sphere.
The projection is performed by dropping a perpendicular to the $z$-axis (see segment OP) and identifying the intersection of this perpendicular line with one of the spheres.  
However, some spins do not have a direct projection onto a sphere. 
For example, in Fig.~\ref{fig5}\textbf{c}, P has no direct projection onto the bottom sphere. 
In such cases, the projected vector fields are complemented by points at the poles of the spheres. 
Specifically, any spin satisfying $m_z \geq \mu_\mathrm{b}$ is mapped to the north pole of the bottom sphere (see P$_\mathrm{b}$). 
Similarly, any spin with $m_z \leq \mu_\mathrm{t}$ is assigned to the south pole of the top sphere (see Supplementary Video~1).  
After projecting, the spins in the upper and lower spheres must be normalized.

The sizes of the top and bottom spheres are determined by the parameters $\mu_\mathrm{t}$ and $\mu_\mathrm{b}$:
\begin{align}
\mu_\text{t} &= + \mathcal{F}(\rho), \nonumber \\
\mu_\text{b} &= - \mathcal{F}(\rho), \nonumber
\end{align}
where 
$\rho$ is the reduced distance, defined as:  
\[
\rho = \sqrt{\left( \frac{2r_x}{L_x} \right)^2 + \left( \frac{2r_y}{L_y} \right)^2}, 
\]
such that $\rho = 0$ at the center of any $xy$-plane and $\rho = 1$ near the boundaries. Here, $L_x$ and $L_y$ represent the dimensions of the simulated domain along the corresponding directions, while $r_x$ and $r_y$ denote the coordinates of a point relative to the domain center.
The function $\mathcal{F}(\rho)$ can be any continuous function satisfying: $\mathcal{F}|_{\rho=0}=0$ and $\mathcal{F}|_{\rho\geq 1}=1$.
A suitable choice is:
\begin{align}
 \mathcal{F}(\rho) = \begin{cases}
\rho^n\quad\text{if}\quad \rho<1, \nonumber \\
1\quad\text{if}\quad \rho\geq 1, \nonumber 
\end{cases} 
\label{f_rho}
\end{align}
where $n$ is a positive integer.
For the textures depicted in Fig.~\ref{fig5}\textbf{d} and \textbf{e} we set $n=4$. 

The explicit equations for the dumbbell map~\cite{VortMerSkyrm} are: 
\begin{equation}
\mathbf{m}_\text{t} = 
\begin{pmatrix}
2{\gamma_\text{t}} m_x  \\
2{\gamma_\text{t}} m_y  \\
- 1 + 2{\gamma_\text{t}}^2(1 + m_z)(1 - \mu_\text{t}) 
\end{pmatrix}, \label{m_t}
\end{equation}
where $$\gamma_\text{t} = 
\begin{cases}
\sqrt{\dfrac{m_z - \mu_\text{t}}{(1 + m_z)(1 - \mu_\text{t})^2}}\quad\text{if}\quad \mu_\text{t}<m_z, \\
0\quad\text{otherwise}, 
\end{cases}$$
and
\begin{equation}
\mathbf{m}_\text{b}  = 
\begin{pmatrix}
2{\gamma_\text{b}} m_x  \\
2{\gamma_\text{b}} m_y  \\
1 - 2{\gamma_\text{b}}^2(1 - m_z)(1 + \mu_\text{b}) 
\end{pmatrix}, \label{m_b}
\end{equation}
where
$$\gamma_\text{b} = 
\begin{cases}
\sqrt{\dfrac{\mu_\text{b} - m_z}{(1 - m_z)(1 + \mu_\text{b})^2}}\quad\text{if}\quad m_z<\mu_\text{b}, \\
0\quad\text{otherwise}. 
\end{cases} \nonumber
$$

\textbf{Calculation of topological indices.} 
To compute the topological indices of an abstract unit vector field~$\mathfrak{m}$ (which may represent $\mathbf{m}$, $\mathbf{m}_\text{t}$ or $\mathbf{m}_\text{b}$, depending on the context), we utilize the winding number integral~\cite{Mapping_degree_theory}, defined as: 
\begin{equation}
W(\mathfrak{m}) = \frac{1}{2 \pi}\int_{\mathbb{S}^1} d{\ell} 
\frac{
\mathfrak{m}_x\,{\partial_\ell}{\mathfrak{m}_y} - \mathfrak{m}_y\,{\partial_\ell}{\mathfrak{m}_x}
}{\mathfrak{m}_x^2 + \mathfrak{m}_y^2}, \label{W}
\end{equation}
along with the Kronecker integral,
\begin{equation}
Q(\mathfrak{m}) = \frac{1}{4\pi}\int_{\mathbb{I}^2} \mathrm{d}r_1\mathrm{d}r_2 \ \mathbf{F}\cdot\hat{\mathbf{e}}_{r_3}, \label{Q}
\end{equation}
and the Whitehead integral:
\begin{equation}
H(\mathfrak{m}) = -\frac{1}{16\pi^2}\int_{\mathbb{I}^2\times\mathbb{S}^1}\mathrm{d}\mathbf{r}\  \mathbf{F}\cdot[(\nabla \times)^{-1}\mathbf{F}], \label{H}
\end{equation}
where $\mathbf{F}$ represents the vector of curvature, defined as
\begin{equation}
\mathbf{F} \equiv 
\left( 
\begin{array}{c}
\mathfrak{m}\cdot[\partial_{r_2}\mathfrak{m}\times\partial_{r_3}\mathfrak{m}]\\ 
\mathfrak{m}\cdot[\partial_{r_3}\mathfrak{m}\times\partial_{r_1}\mathfrak{m}]\\ 
\mathfrak{m}\cdot[\partial_{r_1}\mathfrak{m}\times\partial_{r_2}\mathfrak{m}]   
\end{array}
\right).
\label{F}
\end{equation}
The integration domains in Eqs.~\eqref{W}-\eqref{H} are the elements of~$\Omega$ -- the texture localization domain~\cite{Zheng_23}.
The domain~$\Omega$ is homeomorphic to solid torus, $\mathbb{I}^2\times\mathbb{S}^1$. 
Note that the section of the torus can be chosen up to homeomorphism.
In particular, the square~$\mathbb{I}^2$ can be equivalently replaced by the disk~$\mathbb{D}^2$.
The fields $\mathfrak{m}$ in the above integrals must satisfy the following boundary conditions: they must be $z$-periodic for Eqs.~\eqref{W}–\eqref{H} and must take the value of the base point on the boundary for Eqs.~\eqref{Q} and~\eqref{H}.
The Whitehead integral applicability follows from the existence of natural inclusion $\mathbb{I}^2\times\mathbb{S}^1 \hookrightarrow \mathbb{S}^3$.

The integrals in Eqs.~\eqref{W}, \eqref{Q}, and \eqref{H} can be directly computed when the magnetization vector field is defined analytically.
For vector fields defined on discrete meshes, such as those obtained from numerical simulations, we employ simplex-based equivalents~\cite{Mapping_degree_theory, Hatcher} to compute the corresponding topological indices.
Specifically, for Eqs.~\eqref{W} and~\eqref{Q}, we follow the approach described in Refs.~\cite{Berg_Luscher, VortMerSkyrm}. 
For Eq.~\eqref{H}, we use the numerical integration scheme outlined in Ref.~\cite{Zheng_23}.
In particular, to increase smoothness of the vector field along the $z$-axis, we add a thin transient layer composed of six additional mesh nodes (cuboids) and minimize the Dirichlet energy, ${\int\mathrm{d}\mathbf{r}|\nabla \mathfrak{m}|^2}$, within this layer assuming periodic boundary condition along the $z$-axis.

\textbf{Revealing homotopy types and their groups.}
Our analysis is based on the following commutative diagram — similar to that proposed in Ref.~\cite{VortMerSkyrm} — which relates pairs of topological spaces:
\begin{equation}
\begin{tikzcd}[column sep=large]
(\Omega,\partial\Omega) \arrow{r}{f} \arrow[dashed]{rd}{g}  & (\mathbb{S}^2, X) \arrow{d}{p}  \\
{}   & (\mathbb{S}^2 \vee \mathbb{S}^2, P_0),
\end{tikzcd}
\label{diag}
\end{equation}
where $\partial\Omega$ is the boundary of space $\Omega$ that is homeomorphic to the surface of a solid torus ($\partial\Omega\cong\mathbb{S}^1\times\mathbb{S}^1$), $\mathbb{S}^2$ is the 2-sphere denoting the space of magnetization values, ${X \subset \mathbb{S}^2}$ is a connected subspace of $\mathbb{S}^2$. 
The symbol $\vee$ stands for the wedge sum, and $P_0$ is the common point of the wedge, serving as the base point. 
\CORR{
For clarity, the mathematical symbols and terms used in this section are summarized in Table~\ref{table_glossary}.
}
\begin{table}[h]
\CORR{
\caption{\label{table_glossary}Table of symbols and terms used in this section.}
\begin{ruledtabular}
\begin{tabular}{p{10mm} p{60mm} r}
\multicolumn{1}{l}{Symbol} & \multicolumn{1}{c}{Meaning} & \multicolumn{1}{l}{Ref.} \\
\hline
$\cong$ & homeomorphic (for spaces) or isomorphic (for groups) & \inlinecite{Kosniowski,MathematicsforPhysics} \\
$\times$ & Cartesian product (spaces) or direct product (groups) & \inlinecite{Kosniowski,MathematicsforPhysics} \\
$\backslash$ & set-theoretic difference & \inlinecite{Kosniowski} \\
$\subset$ & subset & \inlinecite{Kosniowski} \\
$\vee$ & wedge sum of pointed spaces & \inlinecite{MathematicsforPhysics} \\
$\rightarrow$ & continuous map (spaces) or homomorphism (groups) & \inlinecite{Kosniowski,MathematicsforPhysics} \\
$\mapsto$ & maps to &  \\
$\hookrightarrow$ & inclusion map & \inlinecite{Kosniowski} \\
$x \circ y$ & composition of maps in the form x(y(...)) & \inlinecite{Kosniowski} \\
$\partial X$ & boundary of a space $X$ & \inlinecite{Kosniowski} \\
$\pi_n(X)$ & $n$-th homotopy group of $X$ & \inlinecite{Kosniowski,MathematicsforPhysics} \\
$\mathrm{Aut}(X)$ & automorphism group of $X$ & \inlinecite{Rotman1995} \\
$\mathrm{GL}_n(\mathbb{Z})$ & group of $n\times n$ integer matrices that have integer inverses & \inlinecite{Rotman1995} \\
$\mathbb{S}^n$ & $n$-dimensional sphere & \inlinecite{Kosniowski} \\
$\mathbb{Z}$ & group of integers under addition & \inlinecite{Rotman1995} \\
$\mathbb{Z}^n$ & free abelian group of rank $n$ & \inlinecite{Rotman1995} \\
\end{tabular}
\end{ruledtabular}
}
\end{table}

Every arrow in the diagram~\eqref{diag} represents a continuous map, and the composition ${g = p\circ f}$ ensures the commutativity of the diagram.
We aim to determine the homotopy types of the map~$f$ and their associated groups. Homotopy types form a group under an operation that governs the combination of topological invariants when magnetic textures merge.

First, consider the case where the space $X$ in diagram~\eqref{diag} is a single point. 
Due to homotopy equivalence, $X$ can also be treated as any subspace of the sphere contractible to a point, such as a spherical cap or a punctured sphere.
In this scenario, the relevant homotopy types form a group, previously determined in Ref.~\cite{Zheng_23}: 
\begin{equation}
G_f \cong \mathbb{Z} \times \mathbb{Z} \equiv \mathbb{Z}^2 \quad (\text{if } X \simeq \text{point}).
\label{Gf_point}
\end{equation}
The classification by the group~\eqref{Gf_point} applies when the magnetic background is sufficiently polarized. 
This includes both the saturated ferromagnetic phase and the conical phase, including cases with distortions, as detailed in Ref.~\cite{Zheng_23}.
The bottom part of the diagram can be disregarded here as it is of little significance for this part of the analysis. 
However, it plays a crucial role in the case discussed below.

Next, consider a more sophisticated and distinct case in which a cone transforms into a helix. In this scenario, the space $X$ is essentially a spherical segment or its homotopy equivalent, $X\simeq\mathbb{S}^1$.
To analyze this case, we begin by examining a non-trivial mapping from the $z$-periodic paths along the boundary of~$\Omega$:
\begin{equation}
\partial\Omega \mathrel{\reflectbox{$\in$}}\  \mathbb{S}^1 \rightarrow X\   \simeq\mathbb{S}^1.
\label{S1toS1}
\end{equation}
The topological invariant of the map~\eqref{S1toS1} is an integer~$v$, representing the winding number that can be computed using Eq.~\eqref{W}, 
$v = W(\mathbf{m})$.  
Physically, $v$ corresponds to the number of periods of helical modulations that fit within the $z$-height of the domain $\Omega$. 
As a result, homotopy types for $f$ can initially be classified into distinct classes, indexed by the integer~$v$.  
To obtain detailed information about such classes, we use the entire diagram~\eqref{diag}.
The auxiliary map~$p$ in the diagram~\eqref{diag} is the surjective dumbbell-like transformation introduced in Ref.~\cite{VortMerSkyrm}.
This transformation gives rise to the emergence of the two unit vector fields $\mathbf{m}_\text{t}$ and $\mathbf{m}_\text{b}$.
Map~$g$ in the diagram~\eqref{diag} is essentially a mapping from a co-H space~\cite{ModernClassicalHomotopyTheory}, so the corresponding to~$g$ homotopy classes form a group.
This group is, in turn, straightforward to derive from the result of Ref.~\cite{Zheng_23}:
\begin{equation}
G_g \cong \pi_2(\mathbb{S}^2 \vee \mathbb{S}^2)\times\pi_3(\mathbb{S}^2 \vee \mathbb{S}^2) 
\cong \mathbb{Z}^2 \times \mathbb{Z}^3 \equiv \mathbb{Z}^5.
\end{equation}
Note that $G_g$ is not our final target group, but rather an intermediate tool for recovering insights into the actual target group $G_f$.
In this regard, it is important that the images of the map~$g$ corresponding to different elements of $G_g$ possess distinct homotopy types for the map~$f$.
To reveal homotopy types in~$f$ (however, generally speaking, not necessarily all of them), we iterate over specific maps for which the induced elements in group~$G_g$ will differ. 
As a result, we reveal the following group homomorphism taking place for every fixed~$v$: 
\begin{align}
\mathbb{Z}^3 &\rightarrow \mathbb{Z}^5\ \cong G_g,\\
(q_\text{t}, q_\text{b}, h) &\mapsto 
(q_\text{t}, q_\text{b}, 
\underbrace{h - v\,q_\text{t}}_{h_\text{t}}, 
\underbrace{h - v\,q_\text{b}}_{h_\text{b}},
-v\,(q_\text{t} + q_\text{b})
).
\label{triplet}
\end{align}
Accordingly, the topological invariants for~$f$ are two 2D indices~\cite{VortMerSkyrm}:
\begin{align}
q_\mathrm{t} &= Q(\mathbf{m}_\mathrm{t}),\\
q_\mathrm{b} &= Q(\mathbf{m}_\mathrm{b}),
\end{align}
that can be computed using Eq.~\eqref{Q}, and one 3D index that can be computed using Eq.~\eqref{H} by one of the following formulas:
\begin{equation}
h = H(\mathbf{m}_\mathrm{t}) + v\,q_\mathrm{t},  
\end{equation}
or
\begin{equation}
h = H(\mathbf{m}_\mathrm{b}) + v\,q_\mathrm{b}.
\end{equation}
Thus, the entire diversity of topological textures embedded in a conical or helical background with~$v$ periods can be classified using an ordered triplet of integers, $(q_{t}, q_{b}, h)$, forming the group $\mathbb{Z}^3\equiv\mathbb{Z}\times\mathbb{Z}\times\mathbb{Z}$ under addition. 
Naturally, the triple has the freedom to be transformed by the action of any matrix in the group of automorphisms, $\mathrm{Aut}(\mathbb{Z}^3)=\mathrm{GL}_3(\mathbb{Z})$.

As a final note, let us emphasize the following. 
From the homotopy point of view, one can continuously transform the conical (helical) phase into a uniformly polarized state.
Such transformation represents a transition from $X\simeq\text{circle}$ to $X\simeq\text{point}$.
This process can equivalently be described as the space~$X$, initially a sphere with two punctures, expanding into a sphere with a single puncture. 
Accordingly, the corresponding inclusions induce the polarization-dependent homomorphisms onto the group~\eqref{Gf_point}:
\begin{align}
\mathbb{S}^2 \backslash \{\hat{\mathbf{e}}_z, -\hat{\mathbf{e}}_z\}  & \hookrightarrow  \mathbb{S}^2 \backslash \{-\hat{\mathbf{e}}_z\}, \\
(q_\text{t},q_\text{b},h) & \mapsto (q_\text{b}, h - v\,q_\text{b}),\label{strong_plus_z}
\end{align}
and 
\begin{align}
\mathbb{S}^2 \backslash \{\hat{\mathbf{e}}_z, -\hat{\mathbf{e}}_z\}  & \hookrightarrow  \mathbb{S}^2 \backslash \{\hat{\mathbf{e}}_z\}, \\
(q_\text{t},q_\text{b},h) & \mapsto (q_\text{t}, h - v\,q_\text{t}).\label{strong_minus_z}
\end{align}
Thus, Eqs.~\eqref{strong_plus_z} and \eqref{strong_minus_z} illustrate how the most general classification presented here, Eq.~\eqref{triplet}, reduces to the specific case discussed in Ref.~\cite{Zheng_23}, where topological classification is done relative to the cone contractible to a point. 
In other words, these equations demonstrate a more universal approach to the topological classification and capture the scenario for transition from $X \simeq \text{circle}$ to $X \simeq \text{point}$.

\CORR{
It is worth noting that alternative attempts to classify similar textures can be found in the literature~\cite{Juha_2009, FosterHarland_2012, knapman2024spacetime, Azhar_2024, tz4v-s5m5, Shaju_2026}.
}

\textbf{Extended Data Video 1.} 
This video shows the \emph{in situ} Lorentz TEM contrast during sequential electric current pulses at a frequency of 1 Hz, resulting in abrupt switching of the magnetic state.

\textbf{Extended Data Video 2.} 
Dumbbell projection from the spin sphere onto a dumbbell composed of top and bottom spheres connected by a handle segment. The map is parameterized by the shape parameters $\mu_\mathrm{t}$, $\mu_\mathrm{b}$.
The black arrow represents an example of a spin being projected. Its projections onto the top and bottom spheres, defined by Eqs.~\eqref{m_t} and \eqref{m_b}, are shown as magenta and yellow arrows, respectively.

\textbf{Data availability}
\\
Data and codes are available upon reasonable request.


\textbf{Acknowledgments}
\\  
X.C. and F.Z. acknowledge financial support from the National Key R\&D Program of China (Grant No. 2024YFA1611100), the National Natural Science Fund for Excellent Young Scientists Fund Program (Overseas) and the General Program (Grant No. 52373226), and the Fundamental Research Funds for the Central Universities.
F.Z. is grateful to the GJYC program of Guangzhou (No. 2024D01J0060), the Guangzhou Basic and Applied Basic Research Foundation (No. SL2024A04J00852), the Guangdong Provincial Quantum Science Strategic Initiative (Grant No. GDZX2401002) and the Xiaomi Young Talents Program.
N.S.K. and S.B. acknowledge financial support from the Deutsche Forschungsgemeinschaft through SPP 2137 ``Skyrmionics'', Grant No.\ KI 2078/1-1 and Grant No.\ BL 444/16-2, respectively.
F.N.R. and O.E. acknowledge support from the Swedish Research Council (Grant No. 2023-04899).
R.E.D-B. is grateful for financial support from the Deutsche Forschungsgemeinschaft (Project-ID 405553726 – TRR~270). 
This project has received funding from the European Research Council under the European Union's Horizon 2020 Research and Innovation Programme (Grant No.~856538 - project ``3D MAGiC'').

\textbf{Author contributions}
\\
F.Z. conceived the project and designed the experiments. 
L.L. prepared the TEM samples. 
X.C. and D.S. performed the experiments and data analysis under the supervision of H.D. and F.Z.
N.S.K. and F.N.R. developed the theory and performed numerical simulations. 
F.N.R. performed homotopy-group analysis.
X.C., F.Z., N.S.K. and F.N.R. prepared the manuscript.
All of the authors discussed the results and contributed to the final manuscript.

\textbf{Competing interests}
\\
The authors declare no competing interests.

\bibliographystyle{sn-nature}
\bibliography{references}

\setcounter{figure}{0}
\renewcommand{\figurename}{\small{\bf Extended Data Fig.}}

\begin{figure*}[htbp]
\centering
\includegraphics[width=1\linewidth]{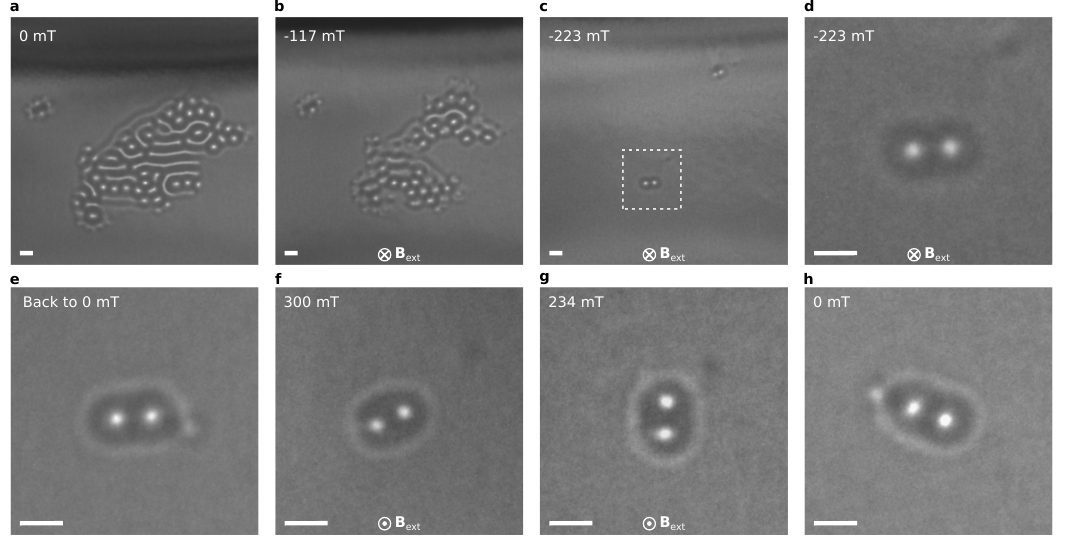}
\caption{\textbf{Hopfion ring nucleation.}
\textbf{a} illustrates the configuration that has been achieved by saturating the sample in the perpendicular magnetic field, subsequently reducing the external field to zero, and finally applying a short electric current pulse.
\textbf{b} and \textbf{c} illustrate the evolution of the contrast under an external magnetic field applied in the negative direction -- opposite to the saturating field. 
Positive and negative field directions correspond to the directions towards the viewer and away from the viewer as indicated by symbols $\otimes$ in \textbf{b}-\textbf{d} and $\odot$ in \textbf{f}-\textbf{g}.
At this step, the large skyrmion clusters collapse, and we \CORR{often} end up with a few configurations with a hopfion ring as in \textbf{c}.
\textbf{d} is a zoomed-in view of the image shown in \textbf{c} (dashed square). 
\textbf{e} the contrast of the hopfion ring around two skyrmion strings after the magnetic field is reduced back to zero.
The images \textbf{f} and \textbf{g} show the evolution of Lorentz TEM contrast as the field increases in the positive direction but nearly the same absolute value as in the negative direction in \textbf{c}. 
\textbf{h} displays the contrast after reducing the external magnetic field to zero again.
It indicates that the hopfion ring remains stable even after a few cycles of applying positive and negative fields.
The images were taken at a defocus distance of 700 $\mu$m and temperature $95$ K.
The scale bar in all images corresponds to 100 nm.
}
\label{fig:S0}
\end{figure*}

\begin{figure*}[htbp]
\centering
\includegraphics[width=1\linewidth]{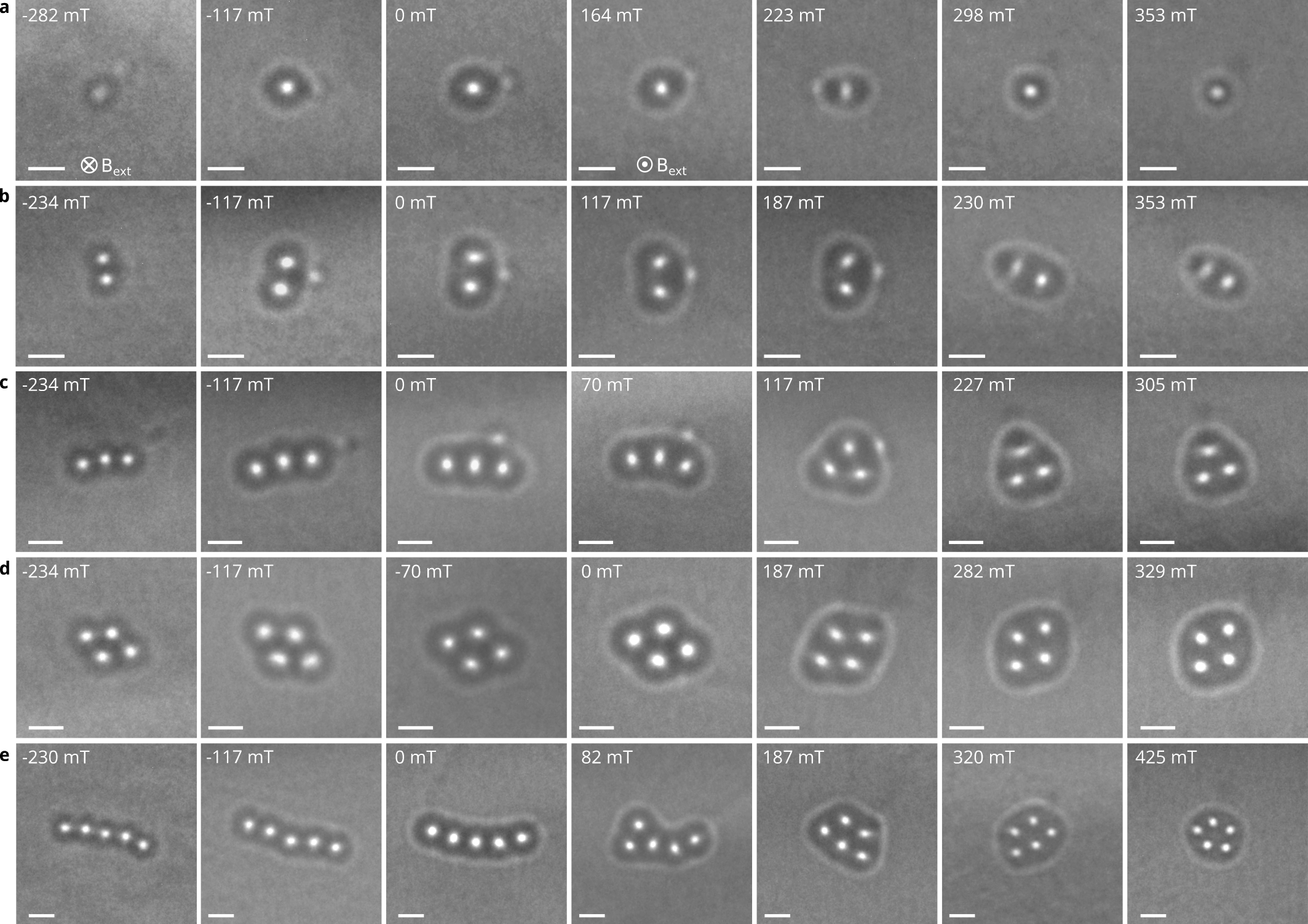}
\caption{\textbf{Field-driven evolution of hopfion rings.} 
Each row of images represents the sequence of over-focus Lorentz TEM images recorded during the magnetic field reversal from a negative to a positive direction.
The strength of magnetic field is labeled in the top-left of each images. 
Positive and negative field directions correspond to fields pointing toward and away from the viewer, as denoted by the symbols $\odot$ and $\otimes$ in \textbf{a}, respectively.
In each row, the negative field decreases from left to right, reaches zero, and then increases in the positive direction.
\textbf{a} and \textbf{b} are the extended versions of Figs. \ref{fig3}\textbf{a} and \textbf{b}, respectively. \textbf{d} the extended version of Fig. \ref{fig2}\textbf{c}.
Under negative fields, bumps are observed in the hopfion rings. As the positive field increases, these bumps gradually diminish and disappear at higher positive fields.
The hopfion ring remains stable within this magnetic field range.
The images were taken at a defocus distance of 700 $\mu$m and temperature $95$ K.
The scale bar in all images corresponds to 100 nm.
}
\label{fig:S8}
\end{figure*}

\begin{figure*}[ht]
    \centering
    \includegraphics[width=1\linewidth]{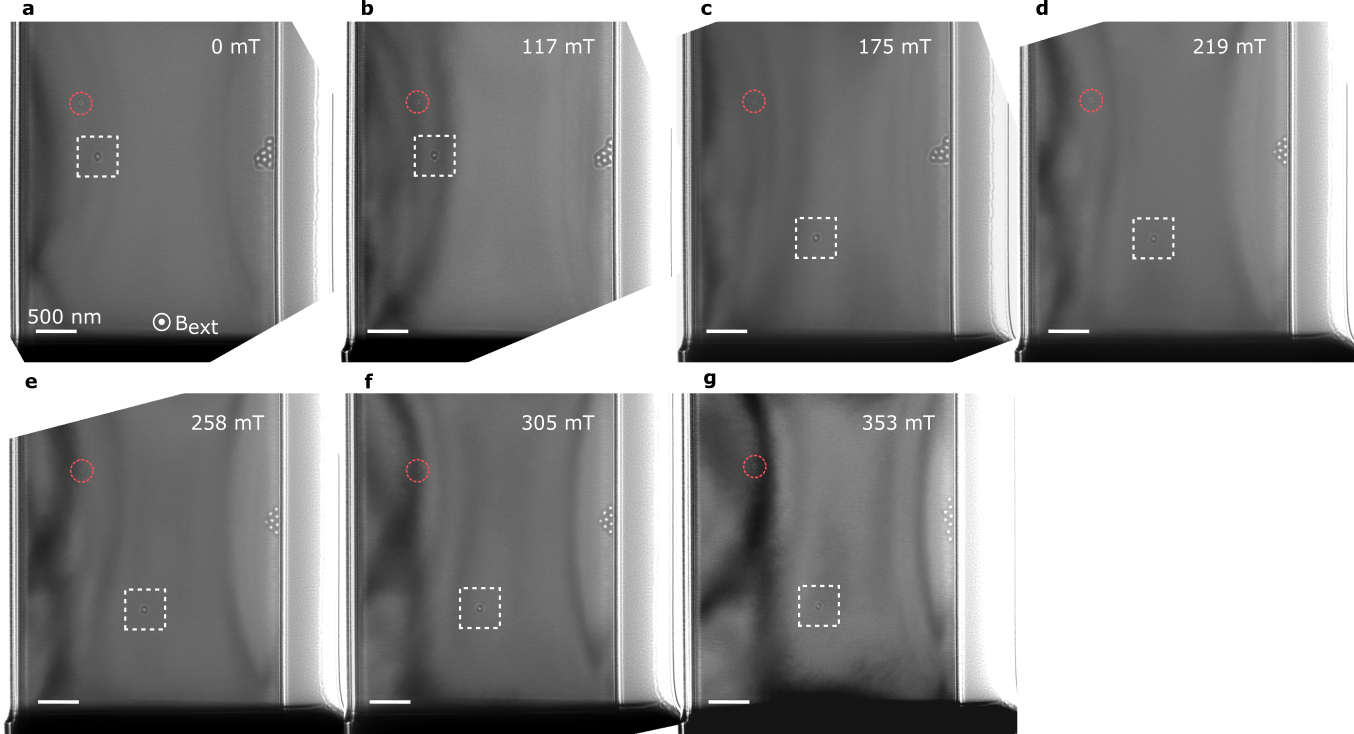}
    \caption{\textbf{Field evolution of the hopfion ring.} An extended version of Fig. ~\ref{fig3}\textbf{a}, showing the entire field of view of the state. 
    Fig.~\ref{fig3}\textbf{a} in the main text displays a zoomed-in region, marked here by a dashed rectangle and rotated clockwise by $90^{\circ}$.
    The texture movement is most pronounced in the field range between 117 mT and 175 mT.
    A red dotted circle marks a surface contamination spot, which serves as a reference point.
    All images were taken in an over-focus Fresnel TEM regime at a defocus distance of 700 $\mu$m and a sample temperature of 95 K. The scale bar in all images is 500~nm.
    }
    \label{fig:S1}
\end{figure*}

\begin{figure*}[ht]
    \centering
    \includegraphics[width=1\linewidth]{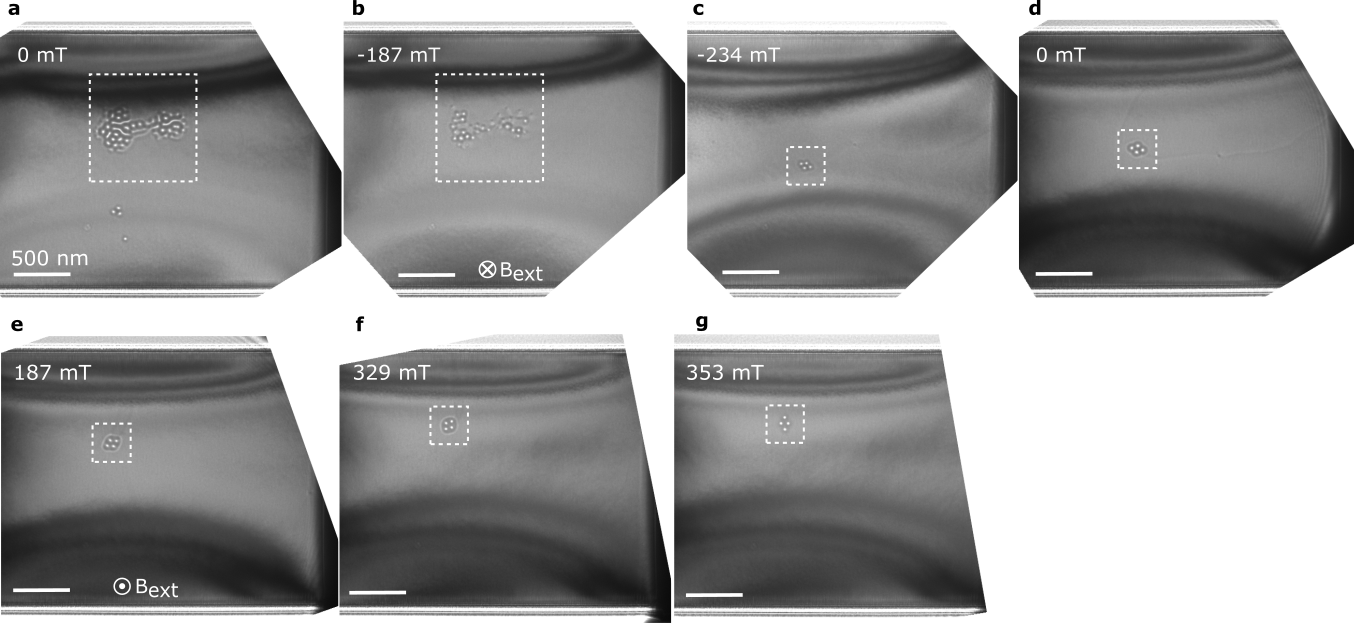}
    \caption{\textbf{Field evolution of the hopfion ring.} An extended version of Fig. ~\ref{fig2}\textbf{c}, showing the entire field of view of the state. 
    The texture moves slightly during field-driven evolution.
    All images were taken in an over-focus Fresnel TEM regime at a defocus distance of 700 $\mu$m and a sample temperature of 95 K.
    The scale bar in all images is 500~nm.
    }
    \label{fig:S6}
\end{figure*}

\begin{figure*}[ht]
    \centering
    \includegraphics[width=1\linewidth]{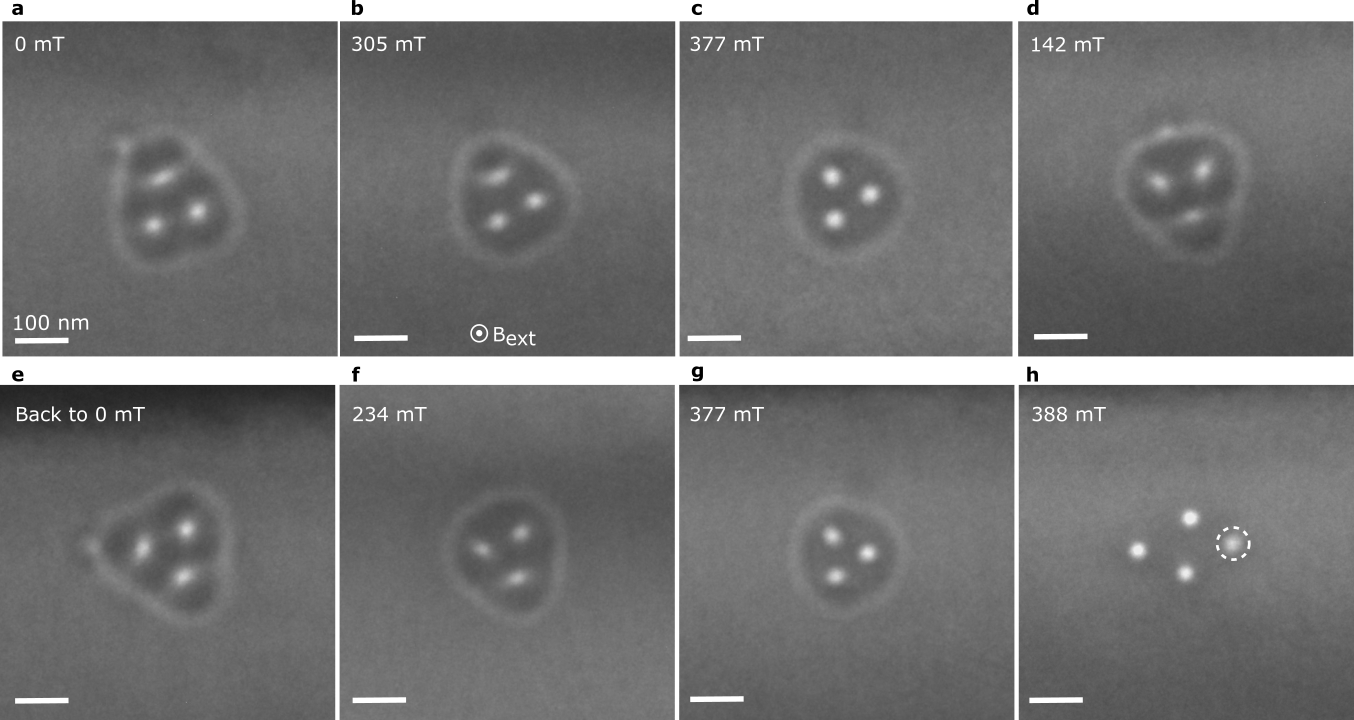}
    \caption{\textbf{Field evolution of the hopfion ring.}
    The images illustrate the emergence of a characteristic bump on the hopfion ring at low magnetic fields and its absence at higher fields.
    The dashed circle marks a low-intercity contrast spot associated with the chiral bobber or dipole string appearing after the collapse of the hopfion ring.
    All images were taken in an over-focus Fresnel TEM regime at a defocus distance of 700 $\mu$m and a sample temperature of 95 K. The scale bar in all images is 100~nm.
    }
    \label{fig:S3}
\end{figure*}

\begin{figure*}[ht]
    \centering
    \includegraphics[width=1\linewidth]{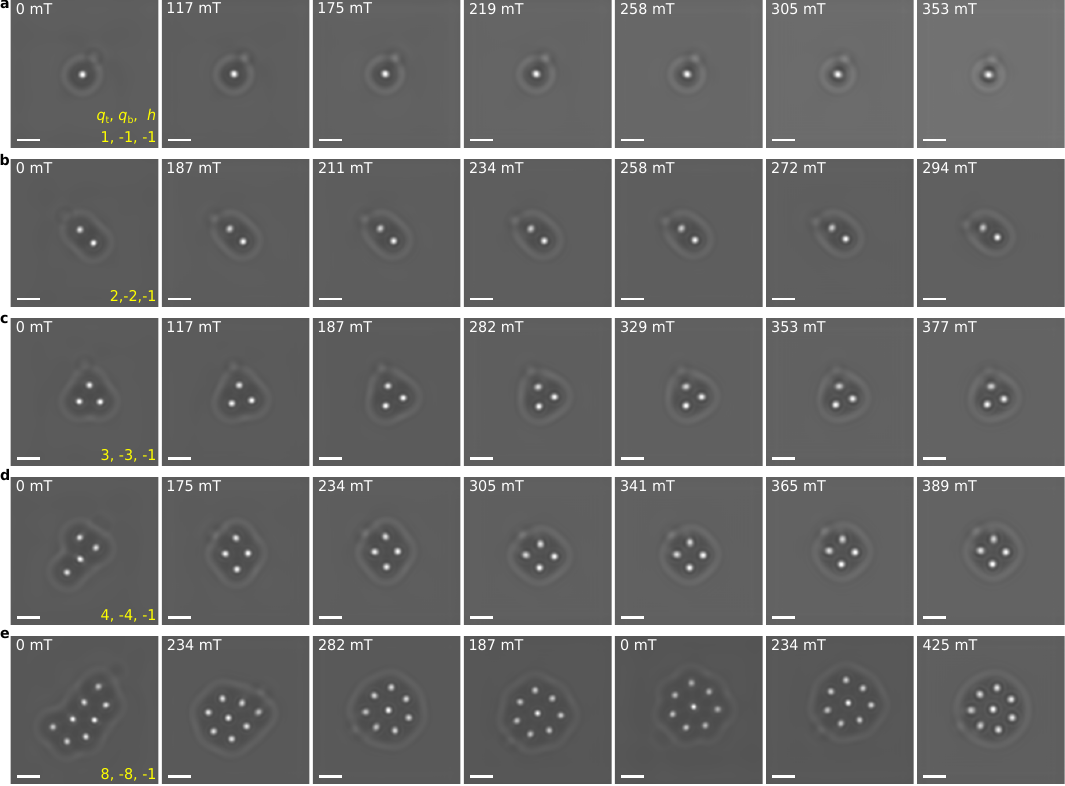}
    \caption{\textbf{Field-driven evolution of hopfion rings in micromagnetic simulations}.
    Each row of images represents the sequence of over-focus Lorentz TEM images calculated for magnetic configurations depicted in Fig.~\ref{fig3} in the main text.
    Corresponding magnetic textures are obtained by energy minimization of the micromagnetic functional.
    The simulated domain has a size of 640 nm $\times$ 640 nm $\times$ 170 nm and periodic boundary conditions in the $xy$-plane (see Method for details).
    All images are calculated assuming a defocus distance of 700~$\mu$m.
    The scale bar in all images is 100~nm.
    }
    \label{fig:SS5}
\end{figure*}

\begin{figure*}[ht]
    \centering
    \includegraphics[width=1\linewidth]{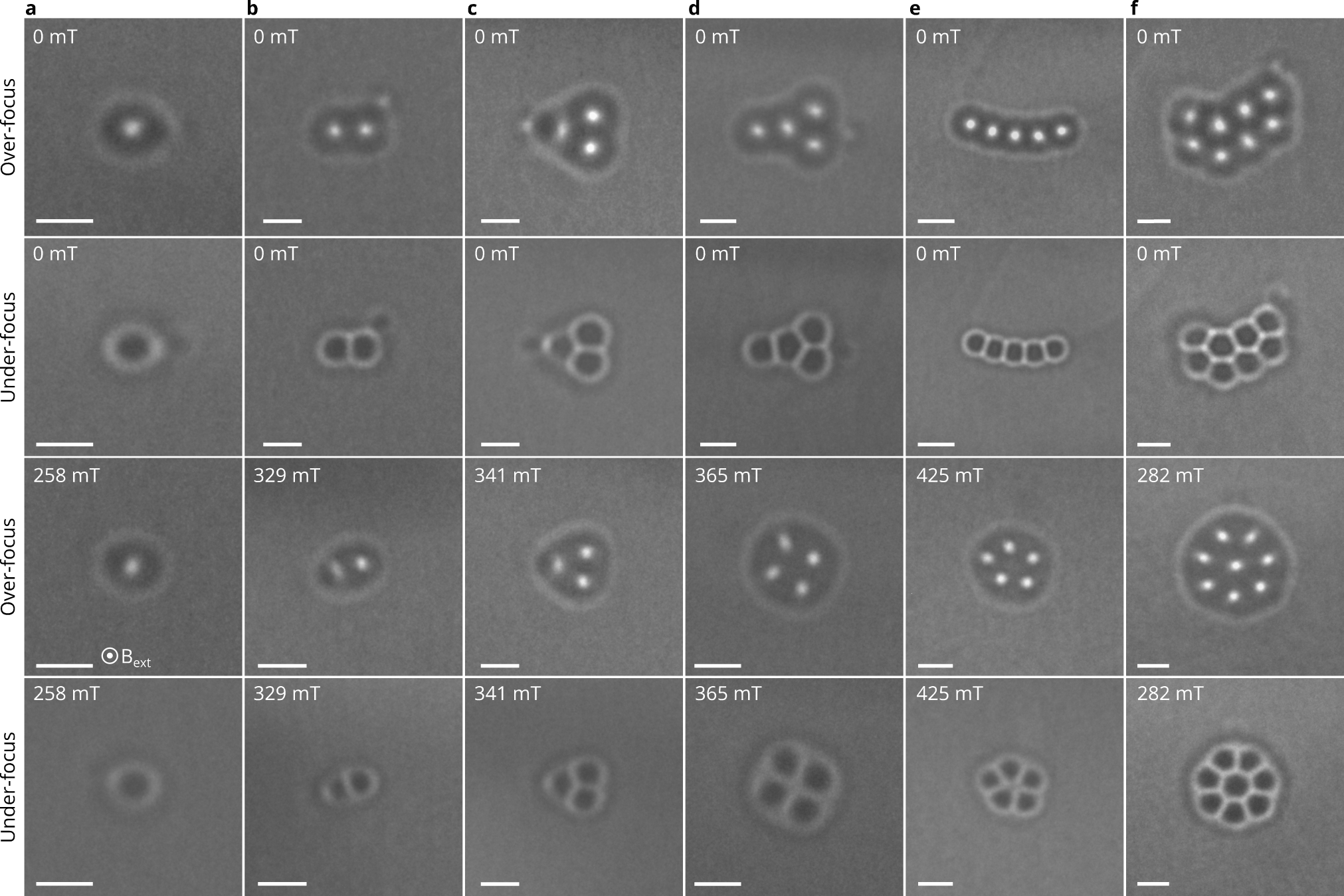}
    \caption{\textbf{Collection of over-focus and under-focus Lorentz TEM images for representative configurations.}
    Each column contains two pairs of Lorentz TEM images acquired in over-focus and under-focus conditions, illustrating the zero-field and high-field states of the same magnetic configuration.
    The magnitude of the applied magnetic field is indicated on the left side of each image.
    All images were recorded at the defocus distance of $\pm$700 $\mu$m, and the sample temperature was 95 K. 
    The scale bar in all images is 100~nm.
    }
    \label{fig:S4}
\end{figure*}

\begin{figure*}[ht]
\centering
\includegraphics[width=1\linewidth]{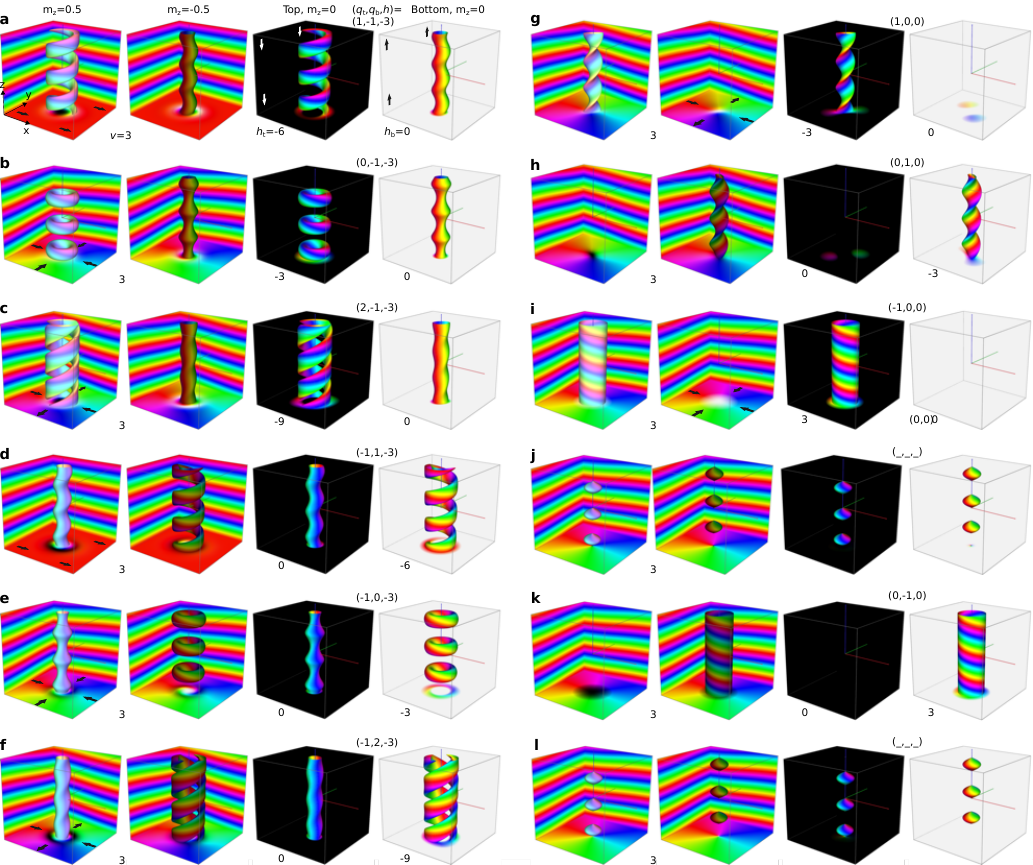}
\caption{\textbf{Collection of topological magnetic textures (skyrmions and vortices).}  
This figure presents a set of magnetic configurations similar to Fig.~\ref{fig5} in the main text and Extended Data Fig.~\ref{fig:E2} but with a different selection of textures. 
Skyrmions in helical backgrounds:  
\textbf{a} Skyrmion in a regular helical vacuum.  
\textbf{b} Skyrmion embedded in a helical vortex.  
\textbf{c} Skyrmion embedded in a helical antivortex.  
\textbf{d}, \textbf{e} and \textbf{f} Same as \textbf{a}, \textbf{b}, and \textbf{c} but with opposite 2D indices.  
Isolated helical vortices and antivortices:  
\textbf{g} Isolated helical antivortex with its core polarity along the positive $z$-axis.  
\textbf{h} Same as \textbf{g} but with opposite core polarity.  
\textbf{i} Isolated helical vortex with its core polarity along the positive $z$-axis (unstable).  
\textbf{j} The relaxed configuration obtained after energy minimization of the state shown in \textbf{i}. Such a configuration is known in the literature as \textit{screw dislocation} of a particular type~\cite{Garst_22}. Since this configuration contains Bloch points (point singularities), the homotopy classification presented here is not applicable.  
\textbf{k}, \textbf{l} Same as \textbf{i} and \textbf{j} but for a helical vortex with opposite core polarity. 
}
\label{fig:E2}
\end{figure*}

\begin{figure*}[ht]
\centering
\includegraphics[width=1\linewidth]{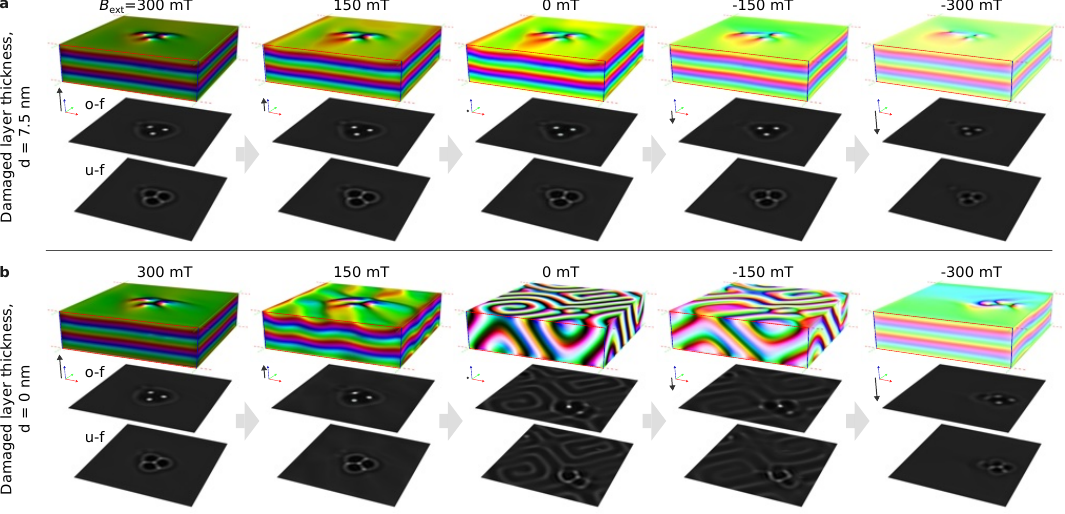}
\caption{\textbf{ Effects of Surface Damaged Layer}. 
The micromagnetic simulations were performed in a domain of size $640 \times 640 \times 170$ nm$^3$ (mesh density $256 \times 256 \times 68$ cuboids) and periodical boundary conditions in the $xy$-plane.
The external magnetic field is applied parallel to the $z$-axis. 
In \textbf{a}, the thickness of the surface damaged layer on both top and bottom surfaces is $7.5$ nm. For the damaged layer properties, see Methods.
In \textbf{b}, we assume there are no damaged layers. 
Each image shows magnetization on the simulated box's surface, represented by the standard color code, and two over-focus (o-f) and under-focus (u-f) Lorentz TEM images on the top and bottom, respectively. 
In both cases, the states are obtained by sequentially reducing the field from 300 mT to -300 mT, as indicated by arrows.
Note the sequence of states in \textbf{a} is fully reversible as in experimental observations, while the states depicted in \textbf{b} are not reversible. 
In the absence of the damaged layer, with a decreasing external magnetic field, the surface modulations penetrate the whole film thickness and form a labyrinth texture.
The configurations depicted in \textbf{a} for three skyrmions and one hopfion ring are identical to those shown in Fig.~\ref{fig2} in the main text.
In contrast, micromagnetic simulations that exclude the damaged layer predict a contrast pattern that is not observed in our experiment. 
}
\label{fig:S7}
\end{figure*}

\begin{figure*}
    \centering
    \includegraphics[width=\linewidth]{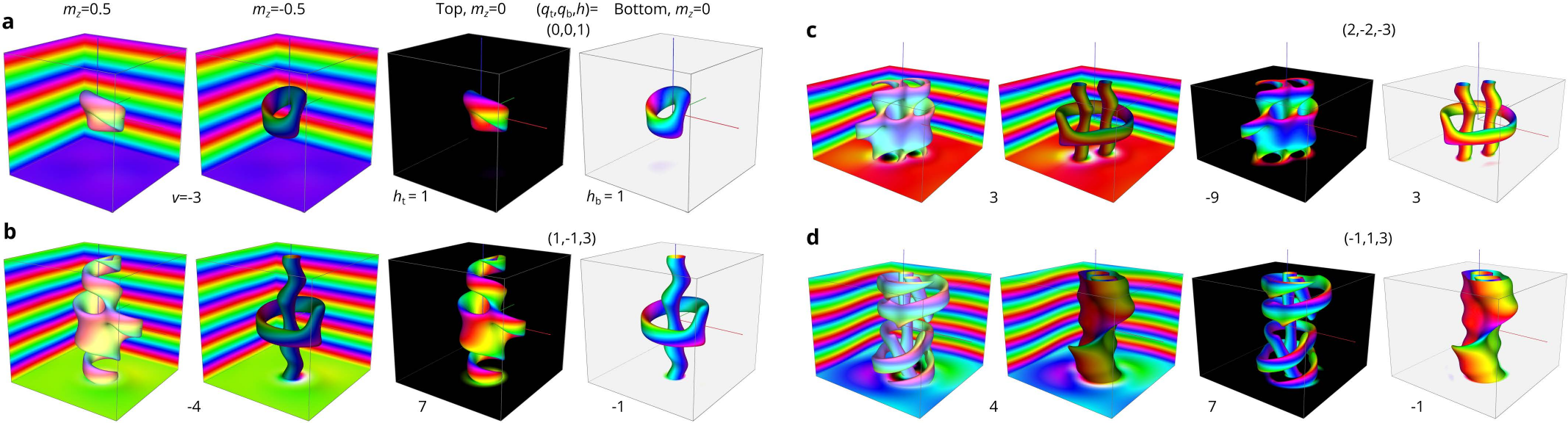}
    \caption{
    \textbf{Collection of topological magnetic textures with different topological index $(q_\mathrm{t}, q_\mathrm{b}, h)$.}  
    \CORR{
    This figure presents a set of magnetic configurations similar to Fig.~\ref{fig5} in the main text and Extended Data Fig.~\ref{fig:E2} but for a different set of textures. 
    \textbf{a} and \textbf{b}, are isolated hopfions embedded into a helix (heliknoton) and hopfion ring that are similar to those in Fig.~\ref{fig5}\textbf{f} and \textbf{g}, respectively, but for the systems with opposite (left-handed) chirality.
    \textbf{c}, Single hopfion ring on a pair of non-braiding skyrmion strings.
    \textbf{d}, A hybrid skyrmion string, as described in Ref.~\cite{Kuchkin22}. 
    }
    }
    \label{fig:E1}
\end{figure*}

\begin{figure*}[ht]
\centering
\includegraphics[width=1\linewidth]{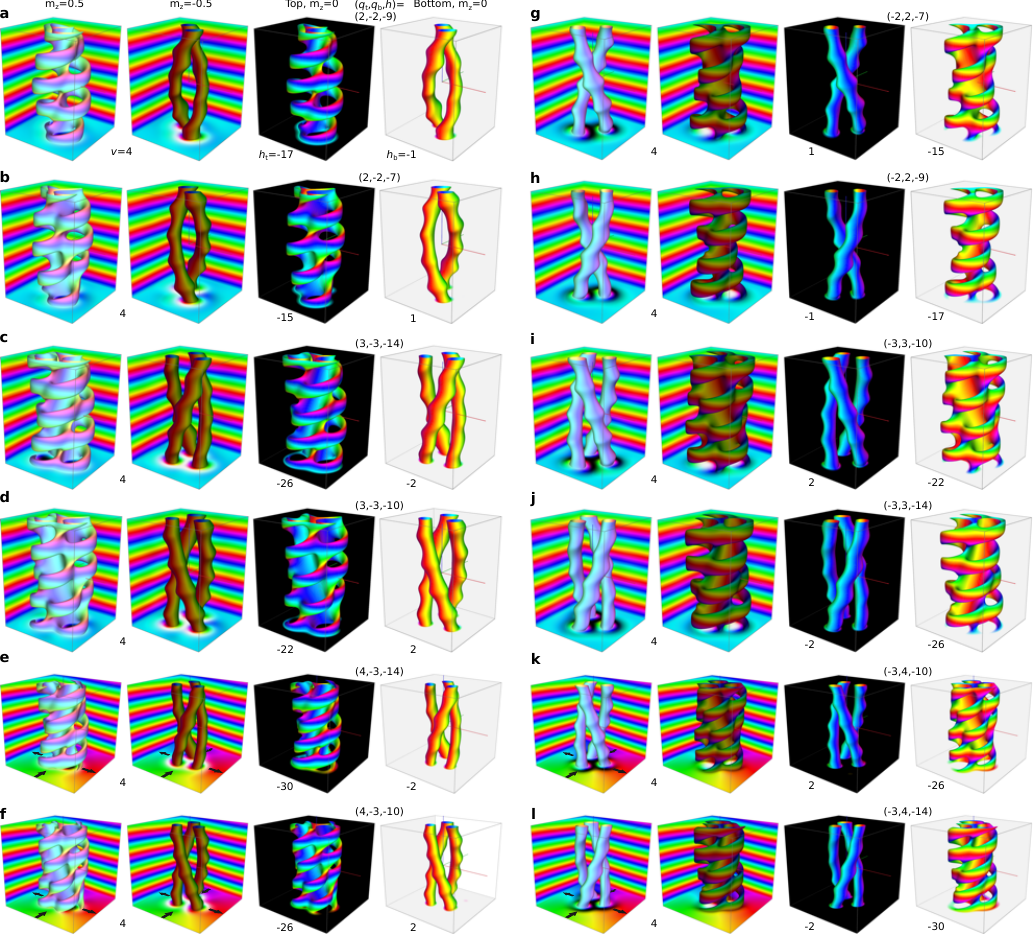}
\caption{\textbf{Collection of topological magnetic textures (skyrmion and antiskyrmion braids).}
This figure presents a set of configurations similar to
\CORR{to Fig.~\ref{fig5} in the main text and }Extended Data Fig.~\ref{fig:E2} but featuring various skyrmion braids with different numbers of skyrmion strings, skyrmion topological indices, and braid twisting angles.
Notably, the skyrmion braids in \textbf{e}, \textbf{f}, \textbf{k}, and \textbf{l} are embedded in a helical antivortex, whereas the remaining configurations reside in a regular helical background.
}
\label{fig:E3}
\end{figure*}

\end{document}